\documentclass{egpubl}
\usepackage{eg2026}
 
\ConferencePaper        % uncomment for (final) Conference Paper
\CGFccby
\usepackage[T1]{fontenc}
\usepackage{dfadobe}  
\usepackage{amssymb}
\usepackage{amsmath}
\usepackage{makecell}
\usepackage{multirow}
\usepackage{booktabs}
\usepackage{array}
\usepackage{utfsym}
\usepackage{color}
\UseRawInputEncoding

\usepackage{cite}  % comment out for biblatex with backend=biber
\BibtexOrBiblatex
\electronicVersion
\PrintedOrElectronic
\usepackage{egweblnk} 
\title[ZeroScene: A Zero-Shot Framework for 3D Scene Generation from a Single Image and Controllable Texture Editing]%
      {\vspace{-25pt} ZeroScene: A Zero-Shot Framework for 3D Scene Generation from a Single Image and Controllable Texture Editing \vspace{-25pt}}

\author[X. Tang, R. Li \& X. Fan]
{
\parbox{\textwidth}{\centering 
X. Tang$^{1,2}$\orcid{0009-0004-8931-4336},
R. Li$^{2}$\orcid{0000-0002-0882-8510}
and X. Fan$^{3,2,4}$\orcid{0000-0002-9660-3636}
}
\\
{\parbox{\textwidth}{\centering 
$^1$Harbin Institute of Technology, Shenzhen, China \\
$^2$Pengcheng Laboratory, China \\
$^3$Harbin Institute of Technology, China \\
$^4$Harbin Institute of Technology, Suzhou Research Institute, China
}
}
\vspace{-8pt}
}

\begin{document}

\teaser{
 \vspace{-0.8cm}
 \includegraphics[width=\linewidth]{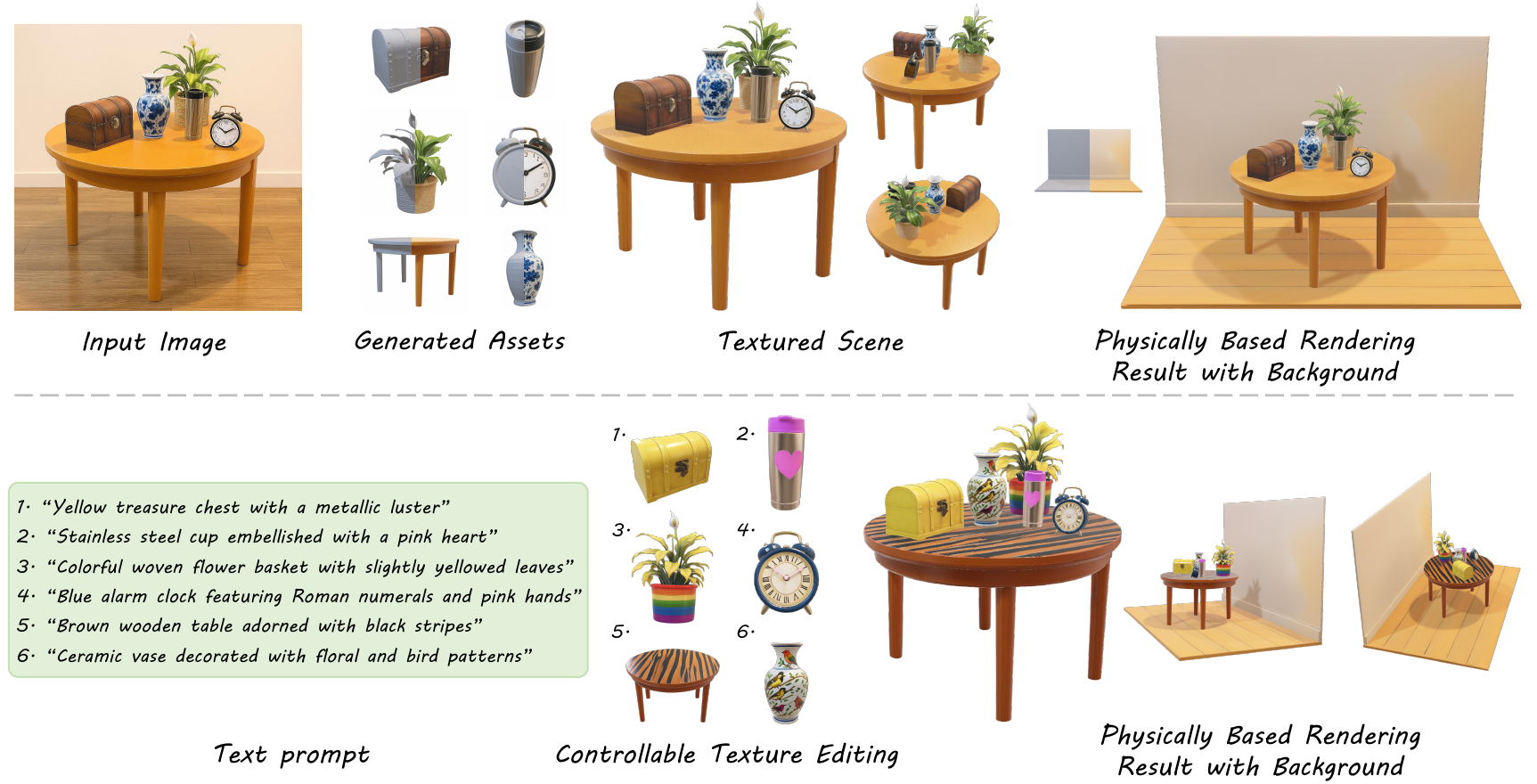}
 \centering
 \vspace{-0.6cm}
 \caption{ZeroScene is capable of generating independent foreground instances while employing dedicated background modeling, thereby reconstructing a complete scene that is spatially precisely aligned with the input image. Moreover, it supports high-fidelity texture editing for objects within the scene and enables physically based rendering.}
\label{fig:teaser}
}

\maketitle

\begin{abstract}
In the field of 3D content generation, single image scene reconstruction methods still struggle to simultaneously ensure the quality of individual assets and the coherence of the overall scene in complex environments, while texture editing techniques often fail to maintain both local continuity and multi-view consistency. In this paper, we propose a novel system \textbf{ZeroScene}, which leverages the prior knowledge of large vision models to accomplish both single image-to-3D scene reconstruction and texture editing in a zero-shot manner. ZeroScene extracts object-level 2D segmentation and depth information from input images to infer spatial relationships within the scene. It then jointly optimizes 3D and 2D projection losses of the point cloud to update object poses for precise scene alignment, ultimately constructing a coherent and complete 3D scene that encompasses both foreground and background. Moreover, ZeroScene supports texture editing of objects in the scene. By imposing constraints on the diffusion model and introducing a mask-guided progressive image generation strategy, we effectively maintain texture consistency across multiple viewpoints and further enhance the realism of rendered results through Physically Based Rendering (PBR) material estimation. Experimental results demonstrate that our framework not only ensures the geometric and appearance accuracy of generated assets, but also faithfully reconstructs scene layouts and produces highly detailed textures that closely align with text prompts. Leveraging generative artificial intelligence, ZeroScene can transform 2D images into diversified 3D worlds with various styles, showing broad application potential in virtual content creation, such as digital twins and immersive game production, while also effectively supports "real-to-sim" transfer in robotics through the generation of highly realistic and trainable simulation environments. Project page: \href{https://xdlbw.github.io/ZeroScene/}{https://xdlbw.github.io/ZeroScene}.



\end{abstract}  

\section{Introduction}

3D generation technology enables the creation of immersive and realistic objects and environments, significantly enhancing the intuitiveness and interactive experience of digital content. As an important research direction in computer graphics, it holds substantial application value in fields such as digital twins, virtual reality, and embodied intelligence. When we focus on the specific task of single RGB image-to-3D generation, although current methods \cite{zhang2024clay, wu2024unique3d, xiang2025structured, li2025triposg} have demonstrated remarkable performance in generating single objects, their quality still declines significantly when handling complex scenes containing multiple objects. This is primarily due to mutual occlusion and overlap between objects in the single view, leading to loss of detail and multi-view inconsistency in the generated assets. Furthermore, the lack of modeling of spatial relationships between objects often results in physically implausible generated scenes, manifested as abnormal object poses, mutual penetration, floating, or failure to form reasonable support relationships under physical contact constraints. Although several recent works have explored compositional scene synthesis, they still commonly suffer from limitations such as insufficient understanding of spatial relationships \cite{han2025reparo, zhou2024zero} and poor geometric and texture quality of generated objects \cite{chen2024comboverse, Ardelean2025Gen3DSR}. Moreover, only a limited number of studies \cite{Ardelean2025Gen3DSR, chen2025physgen3d} have involved the background components within scenes, and exploration in this aspect remains relatively preliminary.

In addition to the paradigm of jointly generating object shapes and textures, another research route focuses on texture editing based on given 3D geometries. Detailed and photorealistic textures are crucial for enhancing the visual appearance of 3D objects, and efficient, high-quality texture synthesis and editing techniques will further advance industries such as animation and gaming. In recent years, with the rapid development of 2D image generation techniques \cite{saharia2022photorealistic, batifol2025flux}, text or image-guided texture generation methods \cite{bensadoun2024meta, xiang2025make} have made significant progress. Many approaches \cite{richardson2023texture, chen2023text2tex, wang2025embodiedgen} leverage pre-trained image diffusion models \cite{rombach2022high, team2024kolors} by incorporating depth constraints to guide the synthesis of RGB images that conform to geometric surfaces. However, these methods often suffer from limited generation quality and multi-view inconsistency, resulting in blurriness or artifacts in the finally synthesized texture. SyncMVD \cite{liu2024text} synchronizes multi-view information at each denoising step and employs cross-view attention to facilitate information sharing, thereby generating detailed and seamless textures. Nevertheless, this approach still struggles to completely avoid the "multi-face janus problem". Therefore, achieving high-quality texture generation that is globally consistent across views, seamlessly aligned, and faithful to user intent remains a significant challenge in current research.

Motivated by these observations, we propose ZeroScene for compositional 3D scene reconstruction from a single RGB image with support for diverse texture editing. For 3D scene generation, we process the foreground and background separately. Initially, we perform foreground instance detection and segmentation on the input image and inpaint the occluded regions to subsequently generate a high-fidelity 3D model for each object. Next, we employ pseudo-stereo techniques to estimate camera parameters and scene depth, from which we extract point cloud representations for both the entire scene and individual instances. Finally, the 3D instances are parameterized, and spatial layout is optimized by minimizing the 3D and 2D projection loss of their point clouds. We also incorporate the real background into the generation process. After masking the foreground instances, the geometric structure is recovered from the complete background point cloud and its spatial parameters are optimized, enabling a unified and harmonious assembly of the entire scene with foreground and background. In the realm of texture editing, we use the normal, position and edge map of a given mesh as conditional constraints. User text prompts are injected into the diffusion process via cross-attention mechanisms to guide the generation of RGB images that conform to the model surface. Subsequently, a mask-guided progressive generation strategy is employed to produce a multi-view consistent set of RGB images. After preprocessing, the final texture map is synthesized through a back projection module. Additionally, we estimate PBR materials of the model to further enhance rendering quality.

We constructed a test set comprising multi-object scene images and diverse 3D mesh models for method validation and performance evaluation. Experiments demonstrate that, even when input images contain object occlusion and complex spatial relationships, ZeroScene not only ensures the generation quality of individual objects but also accurately reconstructs scene layouts, showing significant improvements over existing methods. For texture editing, ZeroScene generates highly detailed and multi-view consistent textures that closely align with text prompts. Moreover, thanks to the incorporation of PBR materials, it significantly enhances rendering realism, outperforming previous approaches. Our main contributions can be summarized as follows: 

\begin{itemize}
\item[$\bullet$] We propose ZeroScene, which is capable of generating multiple high-quality, independent 3D assets from a single image. By comprehending spatial relationships within the scene and jointly optimizing 3D and 2D projection losses of the point cloud for accurate layout recovery, our method constructs a coherent and holistic 3D scene that seamlessly integrates both foreground and background elements.
\item[$\bullet$] By imposing geometric constraints on the diffusion model and introducing a mask-guided progressive image generation strategy, ZeroScene effectively ensures multi-view consistency in 3D texture synthesis, thus enabling diverse and impressive texture editing results for objects in the scene.
\item[$\bullet$] ZeroScene is a highly decoupled zero-shot framework that supports flexible replacement and extension of arbitrary foundational visual model components. All assets generated by this system are represented as explicit triangle meshes, making them readily applicable to various downstream tasks such as geometric editing, game development, and the construction of embodied AI simulation environments.
\end{itemize}
\section{Related Works}

\begin{figure*}[htbp]
    \centering
    \vspace{-0.2cm}
    \includegraphics[width=\textwidth]{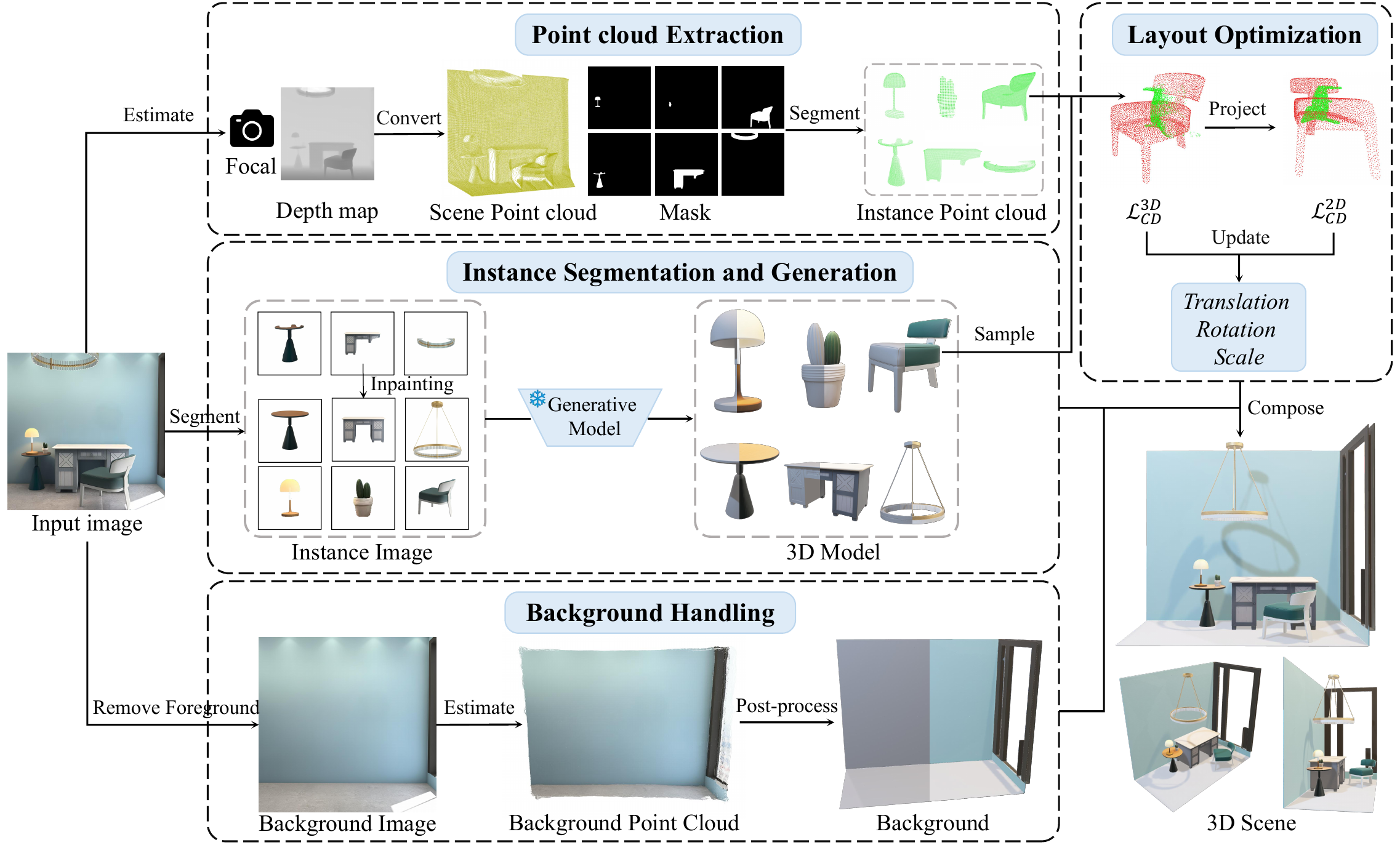}
    \vspace{-0.6cm}
    \caption{\textbf{Overview of 3D Scene Generation.} We decouple the foreground and background of a given image. The assembly of foreground objects is achieved through three steps: instance segmentation and generation, scene point cloud extraction, and layout optimization. For the background environment, we fit planes from point clouds with color information. Finally, the foreground and background are integrated to construct a complete 3D scene that is multi-view consistent and spatially coherent.}
    \vspace{-0.8cm}
    \label{fig:scene generation}
\end{figure*}

\subsection{Single Image Scene Generation}

Generating 3D scenes from a single image is a challenging task, with difficulties including handling object diversity, addressing severe occlusions, and accurately understanding spatial relationships among various parts of the scene. Fortunately, a variety of mature vision models are widely used in related subtasks, such as image segmentation \cite{kirillov2023segment, liu2024grounding, ren2024grounded}, image inpainting \cite{suvorov2022resolution, hurst2024gpt}, 3D reconstruction \cite{wang2024dust3r, yang2025fast3r}, and 3D content generation \cite{ye2025hi3dgen, wu2025direct3d, lai2025hunyuan3d}. Existing methods often rely on these foundational models, integrating decomposition-recomposition \cite{chen2024comboverse, zhou2024zero, daiautomated, han2025reparo, wu2025diorama, Ardelean2025Gen3DSR, chen2025autopartgen}, Large Language Models \cite{liu2024controllable, gu2025artiscene}, scene graph construction \cite{dong2025hiscene}, optimization strategies \cite{chen2024comboverse, han2025reparo, gu2025artiscene, wu2025diorama, yao2025cast}, and point cloud alignment \cite{zhou2024zero, Ardelean2025Gen3DSR} to achieve multi-asset generation and scene layout recovery. Recent studies have attempted to more directly model spatial relationships between objects. For instance, CAST \cite{yao2025cast} proposes a component-aligned generation strategy that aligns content generated in canonical space with scene space by computing model transformation matrices, enabling efficient and accurate 3D scene generation from a single image. MIDI \cite{huang2025midi} captures spatial dependencies during the diffusion process through a multi-instance attention mechanism and Scenegen \cite{meng2025scenegen} employs local-global attention blocks to facilitate interaction among different assets.

Distinct from the component alignment adopted by CAST or the strategy proposed by Zhou et al. \cite{zhou2024zero}, which calculates the loss by randomly sampling discrete points within the contours of original instance masks, we achieve accurate layout recovery by minimizing the joint 3D and 2D projection loss between the predicted instance point clouds and the target point clouds. This effectively rectifies the scale/depth ambiguity inherent in single-view depth estimation, ensuring that the generated models strictly adhere to the spatial layout of the input images. Furthermore, in contrast with methods such as MIDI and Scenegen, which often suffer from object misalignment, structural incompleteness, or redundant artifacts due to the lack of explicit handling of occluded regions, our workflow fundamentally guarantees the structural integrity of occluded objects and the quality of individual assets by separating foreground from background and performing targeted image inpainting prior to 3D lifting.

\subsection{Texture Editing}

Recently, the research focus of AI-based 3D texture generation and editing has gradually shifted from generative adversarial networks \cite{goodfellow2014generative} to diffusion models \cite{ho2020denoising, rombach2022high}. This transition has enhanced the visual diversity and realism of textures, overcoming the limitations in flexibility and efficiency inherent in traditional manual creation methods. Methods based on iterative rendering \cite{richardson2023texture, perla2024easitex, wang2025embodiedgen} leverage depth or normal maps and employ ControlNet \cite{zhang2023adding} to guide diffusion models, effectively aligning textures with geometric models. However, their outputs often suffer from insufficient detail clarity and visual consistency. Mapa \cite{zhang2024mapa} decomposes a 3D mesh into multiple segments and employs segment-controlled strategy to produce 2D images aligned with different mesh components. Several approaches build upon multi-view images \cite{chen2023text2tex, huo2024texgen, zeng2024paint3d, huang2025mv}, incorporating UV inpainting modules to complete missing texture regions and achieve more comprehensive texture synthesis. MVPaint \cite{cheng2025mvpaint} addresses common issues in existing text-to-texture methods, such as local discontinuities, view inconsistency, and over reliance on UV unwrapping quality, through modules for multi-view consistent generation, inpainting of unobserved regions, and UV optimization. Nevertheless, its underlying multi-view diffusion model struggles to comprehend fine-grained information in user inputs, resulting in generated textures that lack detail. The latest research aims to produce textures with physical realism, particularly in terms of light response and material attributes. Paint-it \cite{youwang2024paint} and MaterialMVP \cite{he2025materialmvp} are capable of generating a full set of texture maps, including albedo, roughness, metallic, and normal maps, supporting high resolution PBR pipelines. Material Anything \cite{huang2025material} introduces a confidence mask mechanism to dynamically adjust the influence of illumination, thereby ensuring physical plausibility of material appearance under varying lighting conditions. 

Methods like TEXTure \cite{richardson2023texture} rely excessively on depth-guided image inpainting, which tends to cause seam inconsistencies or visual artifacts when processing the surfaces of complex geometric objects. To address such limitations, we significantly enhance the geometric alignment and visual continuity of textures on complex topological structures, via incorporating richer geometric priors and combining the mask-guided progressive multi-view image generation strategy with a confidence-based back projection module. Moreover, our work not only focuses on the color generation of textures, but also integrates PBR material estimation, thereby endowing the generated textured meshes with favorable physical realism.
\section{Method}

In this section, Sec. \ref{sec:3.1} and Sec. \ref{sec:3.2} will respectively introduce the generation and composition strategies for foreground objects and background environments, which jointly construct a complete 3D scene. In Sec. \ref{sec:3.3}, we describe how ZeroScene performs multi-stylized texture editing on the assets within the scene.

\begin{figure}
    \centering
    %\vspace{-0.3cm}
    \includegraphics[width=\columnwidth]{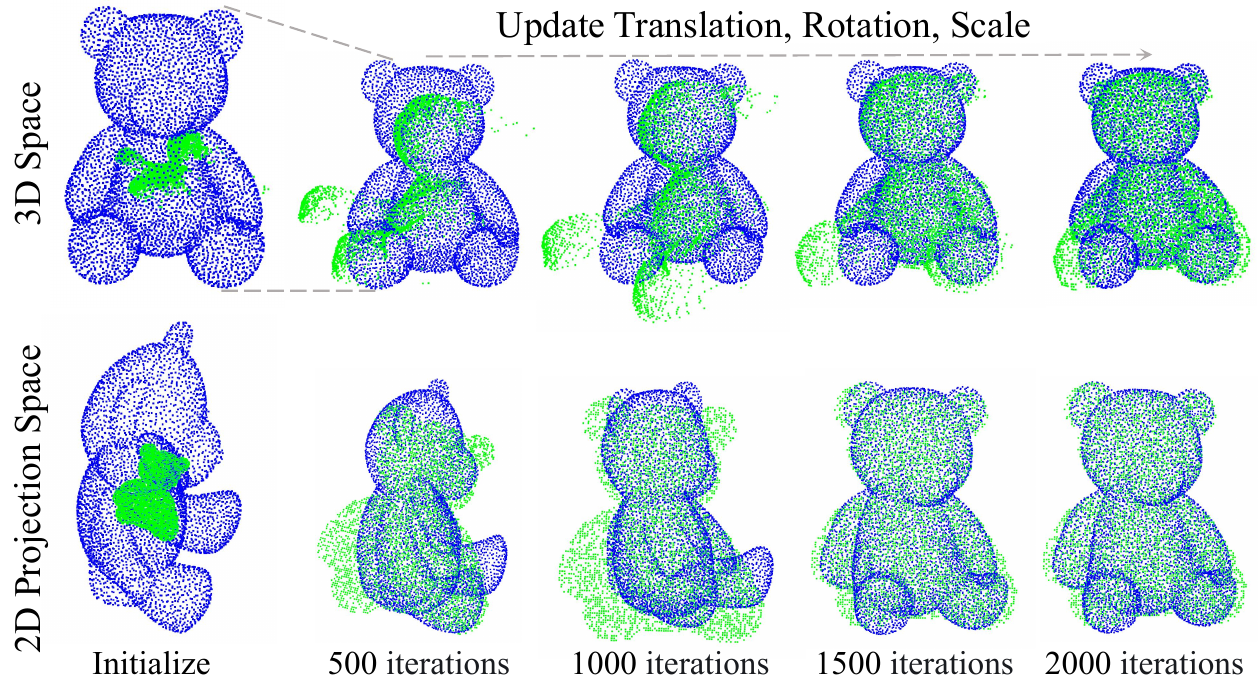}
    %\vspace{-0.7cm}
    \caption{\textbf{Layout optimization process.} Taking the bear doll as an example, the point cloud $\mathcal{M}_i$ of the generated model is depicted in \textcolor{blue}{blue}, i.e., the object to be optimized, while the extracted instance point cloud $\mathcal{PC}_i$ is depicted in \textcolor{green}{green}, indicating the target. We visualize the optimization process in both 3D space and 2D projection space, integrating dual spatial information to achieve superior layout parameters.}
    \vspace{-0.6cm}
    \label{fig:Layout optimization process}
\end{figure}

\subsection{Foreground Object Generation and Composition} \label{sec:3.1}

Accurate reconstruction of a 3D scene from a single image relies on a comprehensive understanding of the geometric structures, appearance characteristics, material properties, and spatial layout of the objects within it. Our core idea is to fully leverage the prior knowledge encoded in pre-trained vision models to construct a robust scene generation framework in a zero-shot manner, as illustrated in Fig. \ref{fig:scene generation}.

\textbf{\textit{Instance Segmentation and Generation.}} Given a single image, our framework first performs foreground-background decoupling, followed by foreground object detection \cite{liu2024grounding}. Using predefined semantic labels, it identifies bounding boxes, category labels, and confidence scores for each instance. Subsequently, based on the detected bounding boxes, a refined instance segmentation module \cite{ren2024grounded} is employed to perform pixel-level extraction on $N$ candidate foreground regions, producing corresponding segmented images and binary masks $\{{p_i, m_i}\}_{i=1}^N$. Due to mutual occlusion between instances in the original image, structural incompleteness often occurs. Directly using such incomplete instances for 3D generation would lead to degraded model quality. To address this, we leverage the powerful semantic reasoning capability of Vision-Language Models (VLMs), e.g., GPT-4o \cite{hurst2024gpt} to locate missing regions via text prompts and generate structurally complete inpainted images. To align with the rapidly advancing 3D generation techniques in the community, our system employs open-source models for image-to-3D conversion, thereby ensuring extensibility and flexible compatibility with future model advancements. Specifically, we use Hunyuan3D 2.5 \cite{lai2025hunyuan3d} to map the inpainted single view instance images into 3D models $\{\mathcal{M}_i\}_{i=1}^N$. The generated 3D assets exhibit accurate geometry, high-fidelity texture consistency with the input image, and are equipped with PBR materials, resulting in high visual realism and rendering quality.

\begin{figure*}
    \centering
    \vspace{-0.2cm}
    \includegraphics[width=\textwidth]{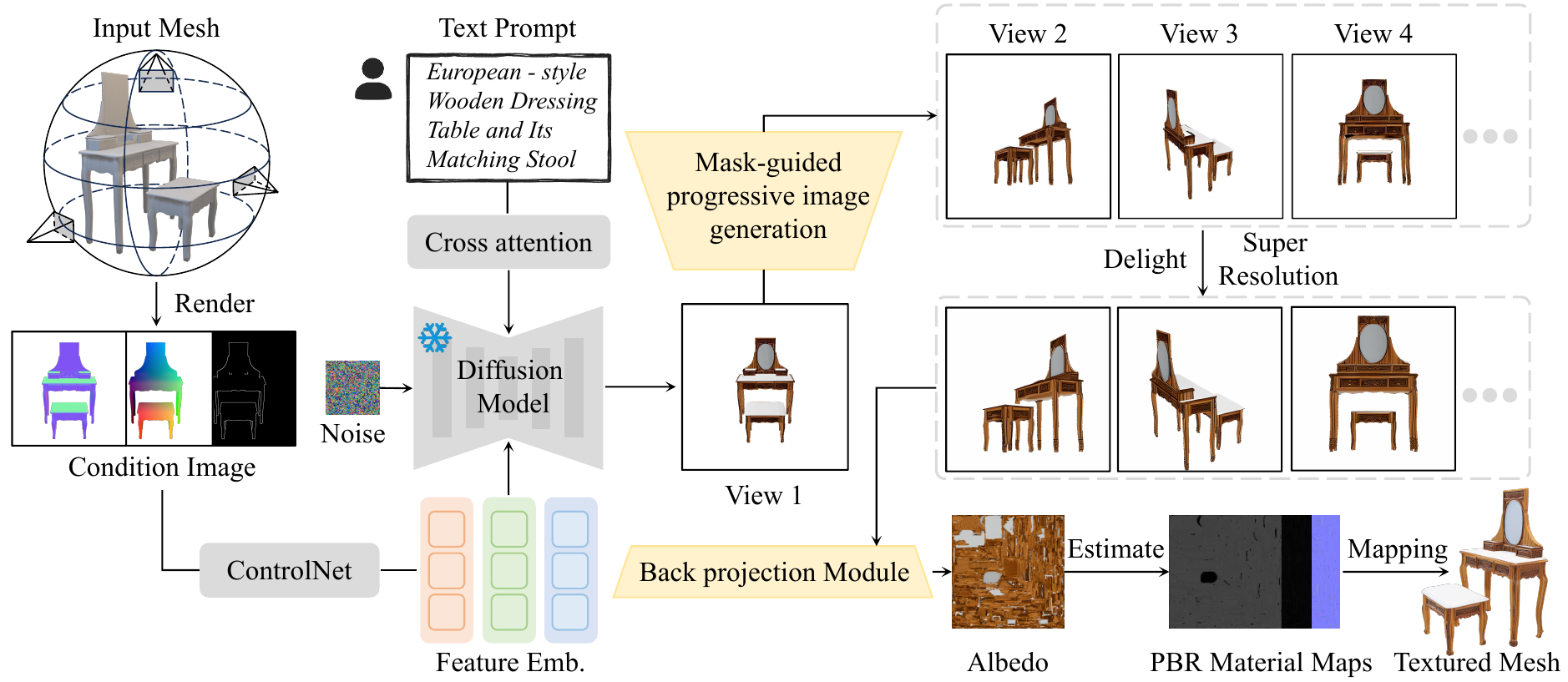}
    \vspace{-0.6cm}
    \caption{\textbf{Overview of Texture Editing.} We utilize generated images for texture synthesis to enable editing. Given a mesh, we render its geometry-aware conditions, which are then injected into a diffusion model along with a text prompt. After obtaining a single-view image aligned with the geometric structure, a mask-guided progressive image generation strategy is employed to synthesize a sequence of RGB images with multiview consistency. The resulting image set is preprocessed with lighting elimination and super-resolution, after which texture is synthesized via a back projection module. Finally, PBR material estimation is incorporated to enhance rendering realism.}
    \vspace{-0.8cm}
    \label{fig:texture editing}
\end{figure*}

\textbf{\textit{Point Cloud Extraction.}} In the second stage, we focus on recovering a 3D point cloud representation of the scene from the images, thereby establishing a geometric foundation for subsequent object layout optimization. DUSt3R \cite{wang2024dust3r} significantly simplifies this process. Trained on a large-scale dataset comprising 8.5 million image pairs, including indoor, outdoor, object-level, and scene-level samples, this model efficiently estimates depth information and camera parameters without relying on camera calibration. It constructs dense mappings between 2D pixels and 3D points, ultimately outputting the 3D structure in the form of pointmaps. Based on the obtained pointmaps, we further extract the complete scene point cloud $\mathcal{PC}$ in the camera coordinate system and its corresponding camera parameters $\mathcal{C}$. Subsequently, we combine the scene point cloud $\mathcal{PC}$ with the instance-level masks $\{{m_i}\}_{i=1}^N$ obtained from the previous stage to perform spatial segmentation, thereby generating independent point cloud representations $\{\mathcal{PC}_i\}_{i=1}^N$ for each of the detected $N$ instances.

\textbf{\textit{Layout Optimization.}}
Finally, ZeroScene strictly adheres to the scene layout of the original image to refine the position and pose of objects in 3D space. Concretely, we reconstruct point clouds $\mathcal{M}$ from the surface of each generated object and parameterize its pose as a set of learnable spatial transformation parameters $\phi = \{T, R, S\}$, denoting translation, rotation, and isotropic scaling, respectively. These parameters are optimized via gradient descent to achieve geometric alignment between the 3D objects and the image scene. The optimization objective is to minimize the 3D Chamfer Distance (CD) loss, which enforces consistency between the instance point cloud $\mathcal{M}_i$ and the target point cloud $\mathcal{PC}_i$ in 3D space. However, relying solely on 3D constraints can lead to instability during the translation and rotation optimization. Inspired by previous studies \cite{chen2021unsupervised, zhou2024zero}, we introduce an additional 2D projection constraint as auxiliary supervision: using known camera parameters $\mathcal{C}$, we project both the instance point cloud $\mathcal{M}_i$ and the target point cloud $\mathcal{PC}_i$ onto the image plane, constructing a 2D CD loss that forces the two projected point sets to match, thereby facilitating 3D structure learning and enhancing cross-dimensional geometric consistency. The overall loss function is defined as:
\begin{equation}
    \mathcal{L} = \lambda_1 \cdot \mathcal{L}_{CD}^{3D}(\mathcal{M}_i, \mathcal{PC}_i) + \lambda_2 \cdot \mathcal{L}_{CD}^{2D}(\text{Proj}_\mathcal{C}(\mathcal{M}_i), \text{Proj}_\mathcal{C}(\mathcal{PC}_i))
    \label{eq:1}
\end{equation}
where $\lambda_1$ and $\lambda_2$ are weighting coefficients balancing the different supervision signals, and $\text{Proj}_\mathcal{C}(\cdot)$ denotes the perspective projection operation based on the camera projection matrix. Fig. \ref{fig:Layout optimization process} illustrates the optimization process of point cloud registration in both 3D and 2D spaces. 

\subsection{Background Handling} \label{sec:3.2}

In scene composition tasks, foreground instances typically occupy the visual focus, while background objects, despite their relatively simple structures, are often overlooked. However, as an essential component of scene construction, the background plays an irreplaceable role: on one hand, the geometric structure of the background provides physical support and serves as the foundation for collision detection in object-scene interaction simulations; on the other hand, as a static contextual environment, it offers the necessary spatial reference and optical basis for global illumination effects such as shadow casting and ambient occlusion. Therefore, our system also incorporates background entities with practical geometric significance (e.g., walls, floors, tables) into the generation scope.

Considering that large portions of the background are frequently occluded by foreground objects, the corresponding point cloud data is incomplete. To address this, we employ a VLM \cite{hurst2024gpt} for iterative object removal, obtaining a complete background image devoid of all foreground instances and their shadows. Based on this image, we utilize DUSt3R \cite{wang2024dust3r} to re-estimate a colored background point cloud. Since the background is generally composed of regular planar surfaces such as floors and walls, we first partition the point cloud into distinct regions according to its normal distribution. For the point cloud subset $\mathcal{B} = \{b_i\}_{i=1}^N$ corresponding to each region, we adopt robust RANSAC sampling combined with the constrained least squares method to extract its principal plane parameters $\mathbf{\pi} = [\mathbf{n}^\top, d]^\top$, where $\mathbf{n} \in \mathbb{R}^3$ denotes the unit normal vector and $d$ is the intercept. Specifically, we optimize the parameters by minimizing the sum of squared distances from the inlier set $\mathcal{S} \subseteq \mathcal{B}$ to the plane, formulated as:
\begin{equation}
    \min_{\mathbf{n}, d} \sum_{b_i \in \mathcal{S}} \|\mathbf{n}^\top b_i + d\|^2, \quad \text{s.t.} \ \|\mathbf{n}\|_2 = 1
    \label{eq:4}
\end{equation}
Subsequently, we perform denoising on the point cloud with the constraints imposed by the fitted plane, and take this as a geometric prior to guide the subsequent Poisson reconstruction, thus yielding a smooth and continuous background mesh with color attributes. Finally, we perform scale calibration on the background model according to the known foreground depth range, and employ Hunyuan3D-Paint 2.1 \cite{feng2025romantex, he2025materialmvp} to generate PBR material textures. Similar to foreground instances, we represent background entities as a set of learnable parameters $\phi = \{T, R, S\}$ in 3D space and optimize their layout using the loss function in Eq. \ref{eq:1}, thereby achieving a unified and harmonious assembly of the complete scene with both foreground and background elements.

\begin{figure}
    \centering
    %\vspace{-0.3cm}
    \includegraphics[width=\columnwidth]{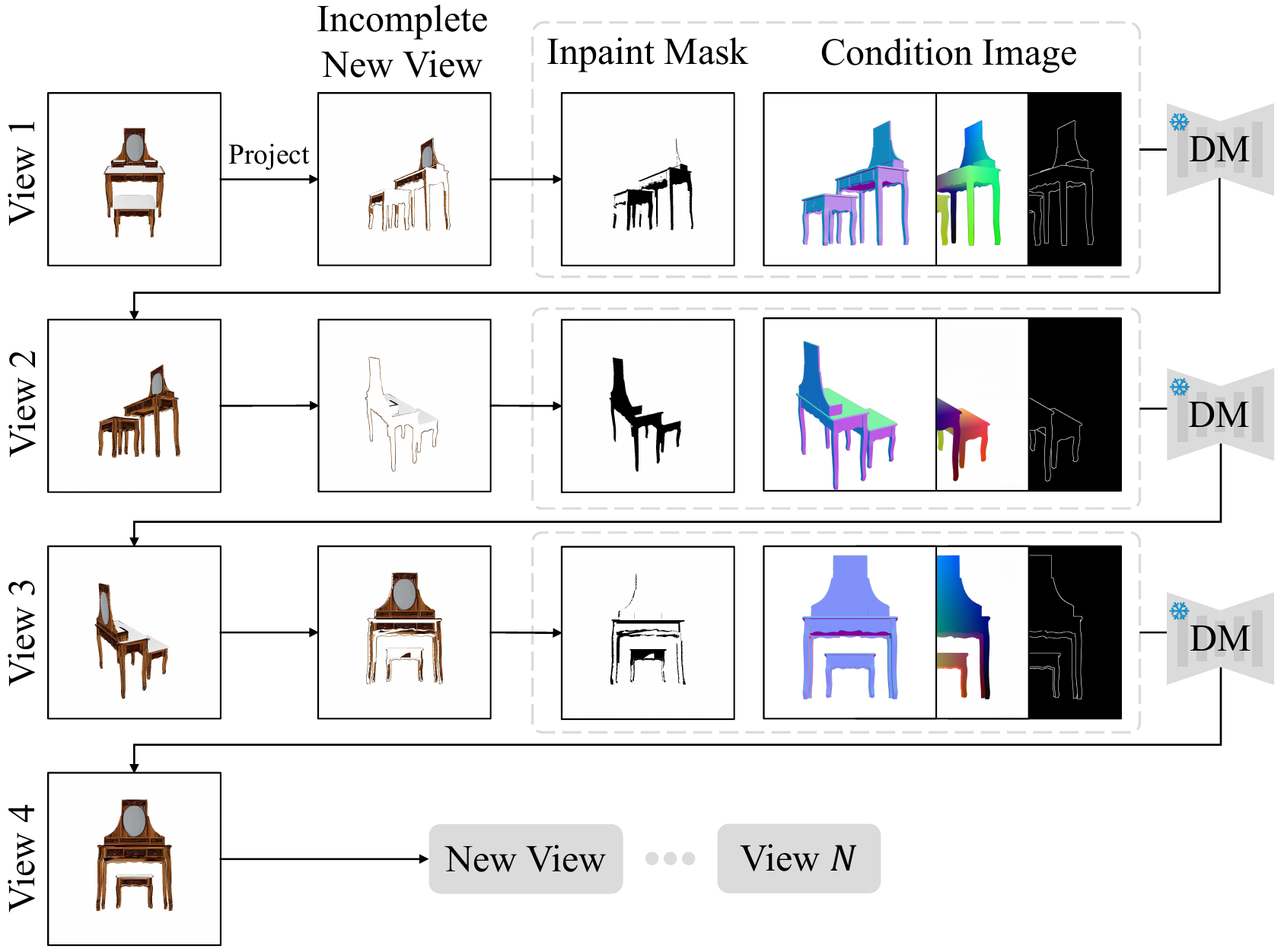}
    \vspace{-0.6cm}
    \caption{\textbf{Mask-guided progressive image generation strategy.} "Project" refers to the process of projecting all currently known regions into the latent space of the next viewpoint. "DM" denotes a pre-trained text-to-image diffusion model \cite{team2024kolors} equipped with ControlNet \cite{zhang2023adding}.}
    \vspace{-0.8cm}
    \label{fig:mask guide generation}
\end{figure}

\subsection{Texture Editing} \label{sec:3.3}

ZeroScene supports multi-stylized texture editing for arbitrary components in a scene, as shown in Fig. \ref{fig:texture editing}. After reconstructing the 3D mesh from the input image (i.e., Sec. \ref{sec:3.1}), our framework is capable of generating diverse and high-quality albedo and PBR material maps based on user-provided text prompts. Our core idea is to synthesize textures through a set of generated RGB images $\{\mathcal{I}_i\}_{i=1}^{N}$ to enable editing. To this end, the set must strictly adhere to the geometric structure of the model while ensuring seamless multi-view consistency.

\begin{figure}
    \centering
    %\vspace{-0.3cm}
    \includegraphics[width=\columnwidth]{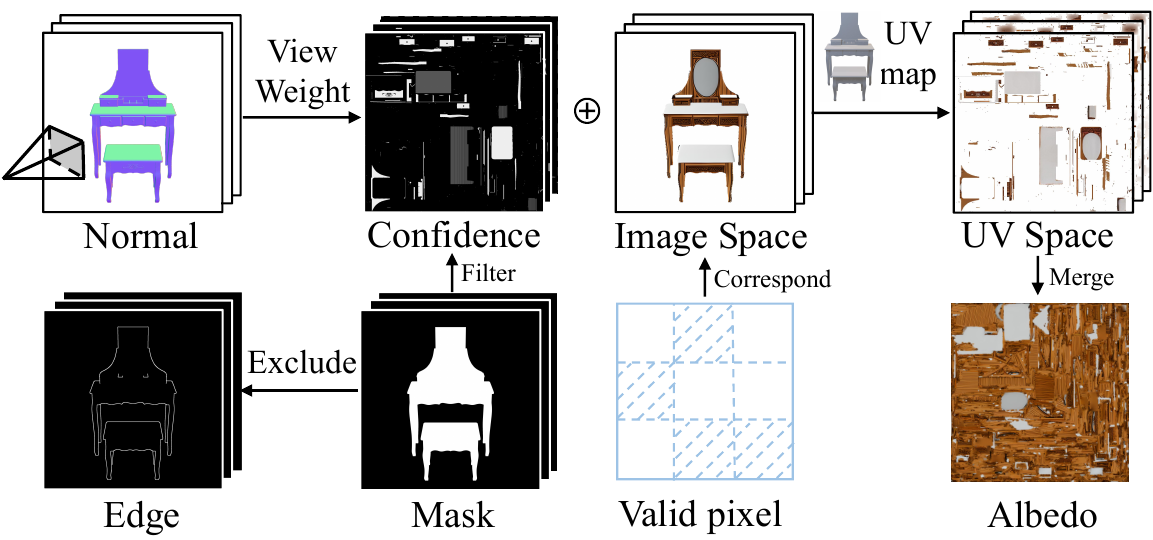}
    \vspace{-0.6cm}
    \caption{\textbf{Back projection module.} The symbol $\oplus$ denotes weighted fusion. We integrate the confidence maps from multiple views with the corresponding color values of valid pixels through weighted fusion. By back projecting multi-view RGB images onto the UV space using the 3D model's UV unwrapping coordinates, we synthesize the albedo map.}
    \vspace{-0.8cm}
    \label{fig:back project}
\end{figure}

To guarantee precise alignment between the generated textures and the underlying geometry, we introduce a geometry-aware module that guides the generation process of the diffusion model. This module enforces texture-to-surface conformity by incorporating spatial constraints and geometric priors. Specifically, we render normal maps, position maps, and depth-based edge detection maps of the input mesh from $N$ predefined camera viewpoints. These 2D information are concatenated into a multi-channel tensor $\mathcal{G} \in \mathbb{R}^{H \times W \times 7}$, which serves as the conditioning input to a ControlNet \cite{zhang2023adding}. Here, the normal map characterizes the surface orientation at each point, ensuring that texture details correspond to the bump variations of the 3D surface. The position map stores the $XYZ$ coordinates of 3D points, acting as spatial anchors to facilitate logical texture distribution across structural partitions. The edge map delineates object boundaries, constrains the valid regions for texture synthesis, and enhances structural articulation along edges. This geometric condition $\mathcal{G}$ is encoded via a branch parallel to the main text-to-image diffusion model backbone \cite{team2024kolors}, and is injected into the denoising process through cross-attention mechanisms to integrate both geometric and textual features, thereby achieving "texture follows geometry" within a single view. The learning objective is defined as:
\begin{equation}
    \mathcal{L}_{\text{DM}} = \mathbb{E}_{\mathbf{z},\mathbf{c},y,\boldsymbol{\epsilon},t} \left\| \boldsymbol{\epsilon} - \boldsymbol{\epsilon}_\theta(\mathbf{z}_t;\mathbf{c},y, t) \right\|_2^2
    \label{eq:2}
\end{equation}
where $\mathbf{z}_t$ denotes the noisy latent representation at timestep $t$, $\mathbf{c}$ represents the geometric condition, and $y$ is the text prompt.

\begin{figure*}[!ht]
    \centering
    %\vspace{-0.3cm}
    \includegraphics[width=\textwidth]{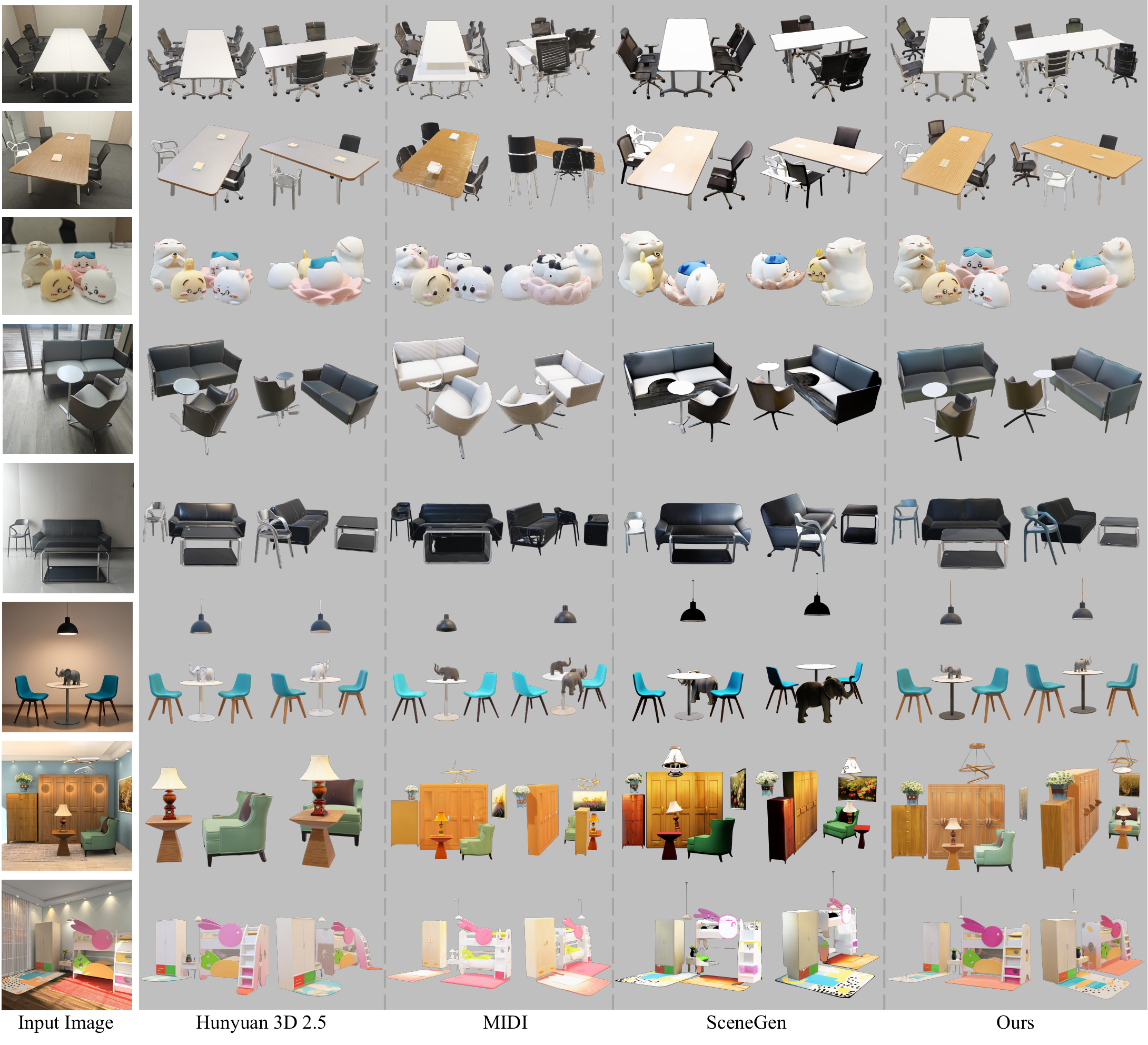}
    \vspace{-0.7cm}
    \caption{\textbf{Qualitative comparisons between ZeroScene and current state-of-the-art single-image scene generation methods.} From left to right: input image, Hunyuan 3D 2.5 \cite{lai2025hunyuan3d}, MIDI \cite{huang2025midi}, SceneGen \cite{meng2025scenegen}, and ZeroScene. The images in the first column from top to bottom are sourced from: real photographs (rows 1-5) and 3D-FRONT \cite{fu20213d} (rows 6-8).}
    \vspace{-0.6cm}
    \label{fig:qualitative comparison1}
\end{figure*}

We observe that when generating only based on the conditional image corresponding to the target new view, it often fails to ensure texture consistency of the overlapping regions between the new image and the existing views, which adversely affects the quality of subsequent texture synthesis. Inspired by relative work \cite{richardson2023texture}, we propose a mask-guided progressive image generation strategy to enhance global consistency across the multi-view image set, as shown in Fig. \ref{fig:mask guide generation}. Concretely, starting from an initial reference view $v_1$, we generate the first complete RGB image $\mathcal{I}_1$ using a geometry-aware diffusion model. For each subsequent target view $v_i$, we first project all previously generated views $\{ \mathcal{I}_j \}_{j=1}^{i-1}$ onto the 2D plane of $v_i$ according to their camera parameters, resulting in an incomplete view containing known texture information. We then generate a binary mask $\hat{m}$ by identifying regions that are visible in the current view $v_i$ but were occluded in all previously generated views. This mask indicates the unknown regions where new texture needs to be synthesized, assigning 0 to these regions and 1 to known areas. Based on this mask, we initialize the latent noise as follows:
\begin{equation}
    z_t = z_t^{\text{proj}} \cdot \hat{m} + z_t^{\text{rand}} \cdot (1 - \hat{m})
    \label{eq:3}
\end{equation}
where $z_t^{\text{proj}}$ denotes the latent representation obtained by encoding the projection of known regions via a Variational Autoencoder, and $z_t^{\text{rand}}$ is randomly initialized noise. The noise map along with the geometric conditioning image of the current view $v_i$, is fed into the diffusion model to perform an inpainting process. This strategy  allows the model to generate semantically plausible and visually coherent textures in the masked unknown regions, while maintaining seamless continuity with known areas and strictly adhering to the geometric constraints of the current view. By iterating this "projection-mask construction-inpainting" pipeline view-by-view, our method facilitates the propagation and expansion of texture information across different views, ultimately yielding a highly consistent multi-view image set $\{ \mathcal{I}_i \}_{i=1}^{N}$.

\begin{table*}[!ht]
    \centering
    \vspace{-0.2cm}
    \caption{\textbf{Quantitative comparisons of scene generation methods.} For geometric evaluation, we employ object-level Chamfer Distance (CD-O) and F-Score (F-S-O), as well as scene-level Chamfer Distance (CD-S) and F-Score (F-S-S). For visual quality assessment, we compute CLIP and DINOv2 image-to-image similarity at both the object level (CLIP-O, DINO-O) and the scene level (CLIP-S, DINO-S). All metrics reported in any table represent average values. Additionally, we report the instance separability and background handling.}
    \setlength{\tabcolsep}{2pt} % 增加列间距，提升可读性
    \renewcommand{\arraystretch}{1.05} % 增加行高，使得表格更通透
    \resizebox{\linewidth}{!}{
    \begin{tabular}{lcccccccccc}
    \toprule[1.5pt]
    \multirow{2}{*}{\centering \makecell{\textbf{Method}}} &  
    \multicolumn{4}{c}{\textbf{Geometric Metrics}} & 
    \multicolumn{4}{c}{\textbf{Visual Metrics}} & 
    \multirow{2}{*}{\centering \makecell{\textbf{Separable} \\ \textbf{Assets}}} &
    \multirow{2}{*}{\centering \makecell{\textbf{Background} \\ \textbf{Handling}}} \\
    \cmidrule(lr){2-5} \cmidrule(lr){6-9}
     & \textbf{CD-O}$\downarrow$ & \textbf{CD-S}$\downarrow$ & \textbf{F-S-O}$\uparrow$ & \textbf{F-S-S}$\uparrow$ &
     \textbf{CLIP-O}$\uparrow$ & \textbf{CLIP-S}$\uparrow$ & \textbf{DINO-O}$\uparrow$ & \textbf{DINO-S}$\uparrow$ & \\
    \midrule
    Hunyuan3D 2.5 \cite{lai2025hunyuan3d} & - & 0.0226 & - & 72.43 & - & 0.876 & - & 0.837 & \usym{2717} & \usym{2717} \\
    MIDI \cite{huang2025midi} & 0.0409 & 0.0384 & 42.76 & 65.58 & 0.814 & 0.841 & 0.782 & 0.819 & \usym{2714} & \usym{2717} \\ 
    SceneGen \cite{meng2025scenegen} & 0.0223 & 0.0416 & 63.63 & 67.17 & 0.835 & 0.827 & 0.846 & 0.812 & \usym{2714} & \usym{2717} \\ 
    \midrule
    Ours & \textbf{0.0163} & \textbf{0.0137} & \textbf{79.35} & \textbf{83.21} & \textbf{0.908} & \textbf{0.913} & \textbf{0.886} & \textbf{0.893} & \usym{2714} & \usym{2714} \\ 
    \bottomrule[1.5pt]
    \end{tabular}
    }
    \vspace{-0.5cm}
    \label{tab:quantitative comparison1}
\end{table*}

\begin{table}
  \centering
  \caption{\textbf{Runtime analysis.}}
  \setlength{\tabcolsep}{2.5pt}
  \resizebox{\linewidth}{!}{
  \begin{tabular}{@{}lccccc@{}}
    \toprule[1.5pt]
    \textbf{Stage} & \textbf{\begin{tabular}[c]{@{}c@{}}Instance Segmentation\\ and Generation\end{tabular}} & \textbf{\begin{tabular}[c]{@{}c@{}}Point Cloud\\ Extraction\end{tabular}} & \textbf{\begin{tabular}[c]{@{}c@{}}Layout\\ Optimization\end{tabular}} & \textbf{\begin{tabular}[c]{@{}c@{}}Background\\ Handling\end{tabular}} & \textbf{Total} \\
    \midrule
    Time & 149.32s & 21.33s & 69.93s & 63.17s & $\sim$300s \\
    \bottomrule[1.5pt]
  \end{tabular}}
  \label{tab:runtime}
  \vspace{-0.6cm}
\end{table}

After acquiring the RGB image set, it must be mapped into UV space for texture synthesis. Prior to this, the raw images undergo two critical preprocessing steps. First, illumination consistency is addressed using a delighting model \cite{hunyuan3d22025tencent} to eliminate inter-view illumination discrepancies caused by the native generation of diffusion models, ensuring overall color coherence. Then, each image is super-resolved \cite{wang2021real} to a resolution of 2K to enhance texture details. As illustrated in Fig. \ref{fig:back project}, to back project the images into UV space, a per-pixel confidence map $\mathcal{C}_i$ is computed for each view $\mathcal{I}_i$, quantifying the reliability of that view contribution to the texture at each point on the model surface. Concretely, we first evaluate the angle between the surface normal and the viewing direction based on the normal map. Surfaces facing the camera directly are assigned higher weights, while pixels with an angle exceeding a predefined threshold $\alpha$ have their confidence set to zero to avoid texture distortion and undersampling caused by excessive grazing angles. To further mitigate potential seams along model contours or depth discontinuous regions, detected edge areas are also excluded from the mask and their confidence is likewise set to zero. This geometric confidence is then multiplied by the prior weight of view $w_i \in \mathcal{W}$ to produce the final confidence map and determine the valid pixels available for texture back-projection. Finally, we traverse all valid pixels in the current view, combine them with the UV unwrapping coordinates of the 3D model, and perform weighted mapping of color values onto the UV space based on confidence. After processing all $N$ views, the final texture map $\mathcal{T}$ is obtained by normalizing the accumulated color values with the total confidence. To further enhance rendering realism, we build upon the generated albedo map and employ Hunyuan3D-Paint 2.1 \cite{feng2025romantex, he2025materialmvp} to estimate metallic, roughness, and bump maps, enabling physically based rendering.
\section{Experiments}

This section begins by introducing the experimental setup in Sec. \ref{sec:4.1}, followed by a comprehensive evaluation of 3D scene generation and texture editing tasks in Sec. \ref{sec:4.2} and \ref{sec:4.3}, respectively.

\subsection{Implementation Details}\label{sec:4.1}

During the layout optimization stage in Sec. \ref{sec:3.1}, the loss weights in Eq. \ref{eq:1} are set as $\lambda_1 = 1$ and $\lambda_2 = 5 \times 10^{-2}$. Each instance is trained for 20 epochs, with 2,000 iterations per epoch. For the first 1,200 iterations, only the 3D CD loss is optimized; in the subsequent 800 iterations, the 2D projected CD loss is additionally incorporated for joint optimization. The epoch with the lowest loss value throughout the training process is selected as the final optimized result. The Adam optimizer is employed, with the learning rate for translation, rotation, and scaling parameters set to 0.01. The aforementioned parameter configuration exhibits excellent generalization ability when processing images from diverse sources, enabling a stable optimization process and accurate estimation of layout parameters. In Sec. \ref{sec:3.3}, we predefine $N = 10$ camera views, including six principal views (weight $\mathcal{W} = 1$) and four oblique views (elevations $\in \{-20^\circ, 20^\circ\}$, azimuths $\in \{45^\circ, 135^\circ, 225^\circ, 325^\circ\}$, weight $\mathcal{W} = 0.1$). The angular threshold is set to $\alpha = 60^\circ$. All experiments are conducted on a single NVIDIA A100 GPU with 40GB of memory.

\subsection{3D Scene Generation} \label{sec:4.2}

\textbf{\textit{Baseline \& Benchmark.}} We compare ZeroScene with several representative single image 3D scene generation methods, including Hunyuan3D 2.5 \cite{lai2025hunyuan3d}, MIDI \cite{huang2025midi}, and SceneGen \cite{meng2025scenegen}. All comparative methods were implemented using their publicly available source code and pre-trained models. The evaluation was conducted on an image set containing multiple foreground objects with significant occlusion relationships, which comprises real photographs, content generated via VLM \cite{hurst2024gpt}, and the publicly available indoor synthetic scene benchmark 3D-FRONT \cite{fu20213d}.

\begin{figure*}
    \centering
    %\vspace{-0.3cm}
    \includegraphics[width=\textwidth]{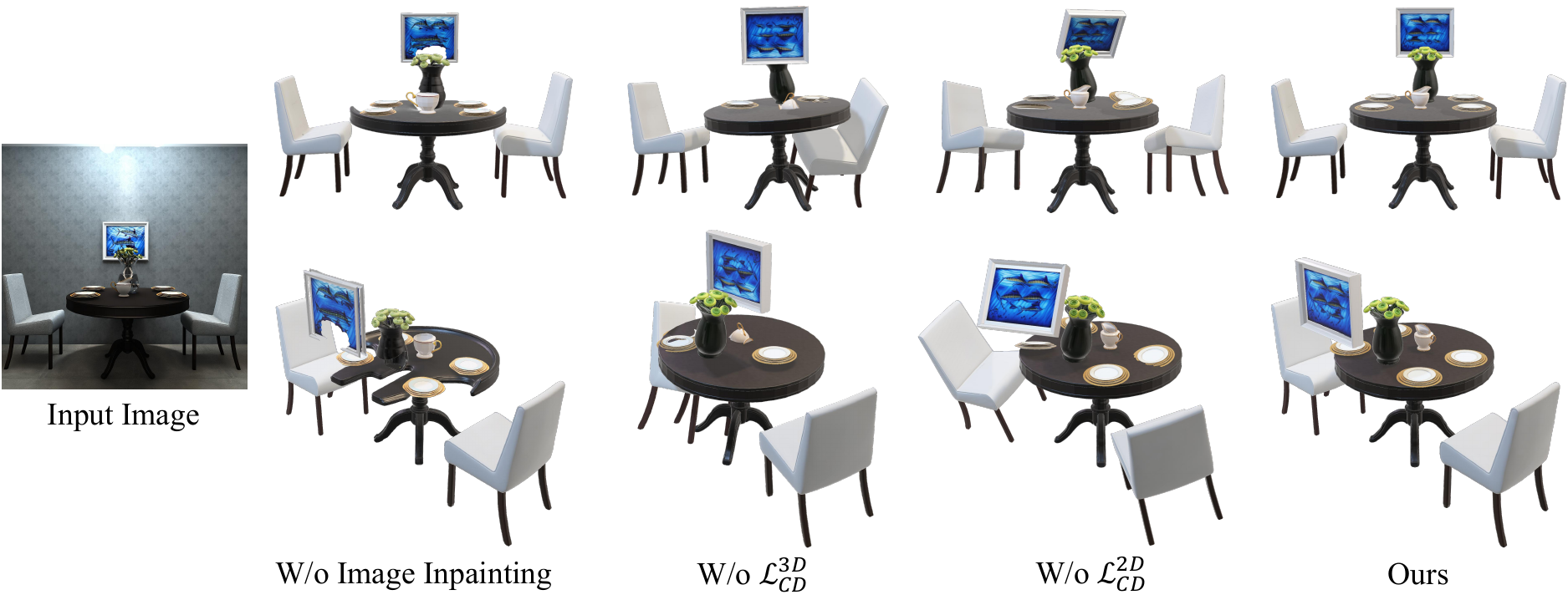}
    %\vspace{-0.7cm}
    \caption{\textbf{Ablation study.} The complete framework ensures the quality of individual assets while accurately recovering the spatial layout, thereby achieving compositional scene generation that is highly aligned with the input and exhibits consistency in both geometry and texture.}
    \vspace{-0.6cm}
    \label{fig:ablation qualitative1}
\end{figure*}

\begin{figure*}[ht]
    \centering
    %\vspace{-0.3cm}
    \includegraphics[width=\textwidth]{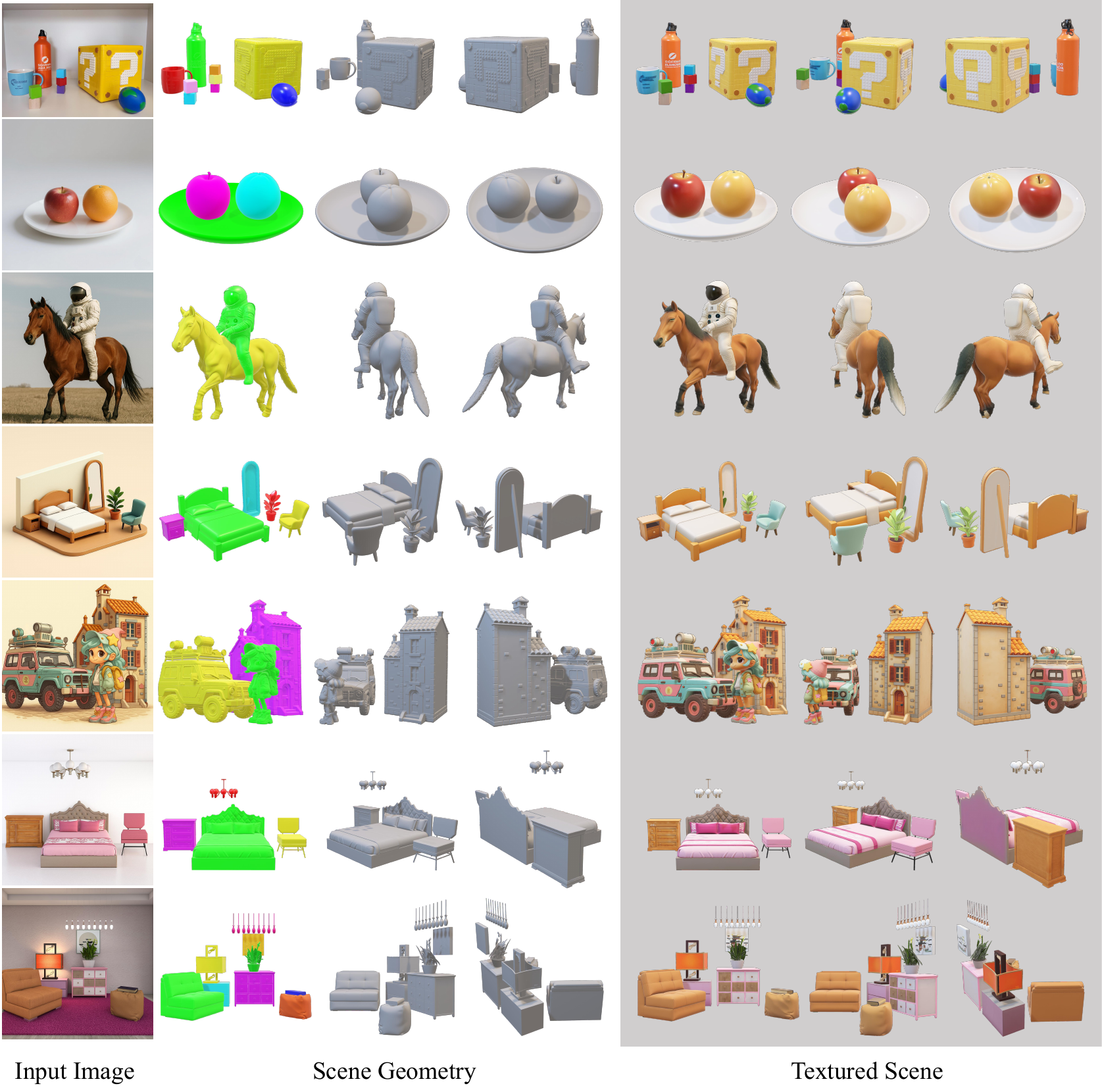}
    %\vspace{-0.7cm}
    \caption{\textbf{More scenes generated by ZeroScene.} From left to right: input image, scene geometry (in the first column, we use distinct colors to annotate objects, indicating they are independent assets), and textured scene. The images in the first column from top to bottom are sourced from: real photographs (row 1), VLM generated images (rows 2-5), 3D-FRONT \cite{fu20213d} (rows 6-7).}
    \vspace{-0.6cm}
    \label{fig:results}
\end{figure*}

\begin{figure*}
    \centering
    %\vspace{-0.3cm}
    \includegraphics[width=\textwidth]{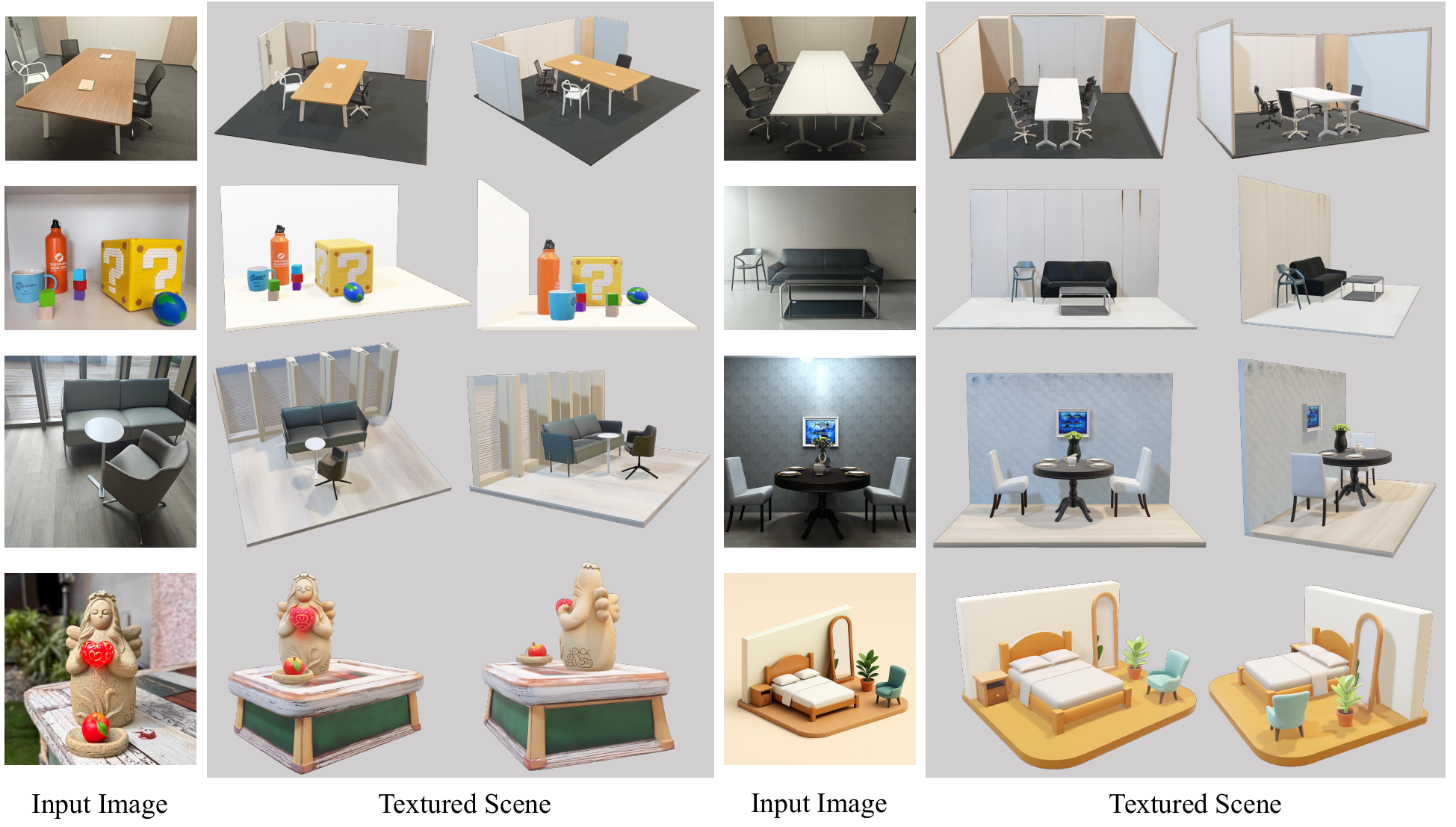}
    %\vspace{-0.7cm}
    \caption{\textbf{More complete scenes including backgrounds generated by ZeroScene.}}
    %\vspace{-0.6cm}
    \label{fig:more results with bg}
\end{figure*}

\noindent\textbf{\textit{Qualitative comparison.}} Since the selected baseline methods lack background generation capabilities, for a fair comparison, Fig. \ref{fig:qualitative comparison1} presents exclusively on the foreground objects generation results of each approach under the same scene input. In terms of geometry and texture quality, Hunyuan 3D 2.5 may exhibit partial geometry asset loss and texture estimation bias in certain scenarios. When the input image contains objects in close proximity or with inclusion relationships, MIDI tends to generate repetitive structures. SceneGen, which employs Trellis \cite{xiang2025structured} as its 3D backbone, suffers from illumination estimation errors in unseen viewpoints during the generation process, leading to localized texture overexposure or underexposure. Additionally, due to the lack of completion processing for segmented instance images, this method may produce erroneous redundant structures when handling occluded objects. In contrast, our method excels in preserving structural integrity of individual objects and high-frequency texture details, while also enhancing rendering realism through PBR materials. Regarding layout construction, Hunyuan 3D 2.5 natively captures image features and produces spatially reasonable asset distributions. However, the multi-instance attention mechanism adopted by MIDI struggles to accurately model spatial relationships between objects in complex scenes. SceneGen, on the other hand, exhibits noticeable object misalignment. Our method achieves accurate reasoning of spatial relationships between objects and ensures multi-view consistency across the entire scene. Additional qualitative results (including background) generated by ZeroScene can be found in Fig. \ref{fig:results} and \ref{fig:more results with bg}.

\begin{table}
    \centering
    \caption{\textbf{Quantitative analysis of ablation study.} "Inp." indicates image inpainting.}
    \setlength{\tabcolsep}{2pt} % 增加列间距，提升可读性
    \renewcommand{\arraystretch}{1.3} % 增加行高，使得表格更通透
    \resizebox{\linewidth}{!}{
    \begin{tabular}{lcccccccc}
    \toprule[1.5pt]
    \multirow{2}{*}{\centering \makecell{\textbf{Ablation}}} &  
    \multicolumn{4}{c}{\textbf{Geometric Metrics}} & 
    \multicolumn{4}{c}{\textbf{Visual Metrics}} \\
    \cmidrule(lr){2-5} \cmidrule(lr){6-9}
     & \textbf{CD-O}$\downarrow$ & \textbf{CD-S}$\downarrow$ & \textbf{F-S-O}$\uparrow$ & \textbf{F-S-S}$\uparrow$ &
     \textbf{CLIP-O}$\uparrow$ & \textbf{CLIP-S}$\uparrow$ & \textbf{DINO-O}$\uparrow$ & \textbf{DINO-S}$\uparrow$ \\
    \midrule
    W/o Inp. & 0.0562 & 0.0514 & 57.50 & 59.83 & 0.774 & 0.861 & 0.703 & 0.825 \\
    W/o $\mathcal{L}_{CD}^{3D}$  & \textbf{0.0163} & 0.0429 & \textbf{79.35} & 63.64 & \textbf{0.908} & 0.793 & \textbf{0.886} & 0.784 \\ 
    W/o $\mathcal{L}_{CD}^{2D}$  & \textbf{0.0163} & 0.0251 & \textbf{79.35} & 70.19 & \textbf{0.908} & 0.836 & \textbf{0.886} & 0.819 \\
    \midrule
    Ours & \textbf{0.0163} & \textbf{0.0137} & \textbf{79.35} & \textbf{83.21} & \textbf{0.908} & \textbf{0.913} & \textbf{0.886} & \textbf{0.893} \\
    \bottomrule[1.5pt]
    \end{tabular}
    }
    %\vspace{-4pt}
    \label{tab:ablation quantitative1}
    \vspace{-0.6cm}
\end{table}

\noindent\textbf{\textit{Quantitative comparison.}} We evaluate the generated 3D scenes from both geometric structure and visual quality perspectives. In terms of geometric structure, we sample discrete points from the surfaces of both the generated assets and the ground truth models (i.e., 3D-FRONT \cite{fu20213d}), and align the reconstructed point cloud with the ground truth point cloud. Subsequently, common geometry evaluation metrics, including Chamfer Distance and F-Score, are computed at both the object level and the scene level. For visual quality, we render the full generated scene as well as individual instances based on the camera parameters of the input images, using the input scene images and instance segmentation images as ground truth. CLIP \cite{radford2021learning} similarity  and DINOv2 \cite{oquab2024dinov2} similarity are adopted as metrics to quantify the texture quality of the generated assets.

As shown in Tab. \ref{tab:quantitative comparison1}, ZeroScene consistently outperforms existing methods across both geometric and visual dimensions, at both the object and scene levels. This advantage stems from its ability to effectively capture local asset details and accurately recover scene layouts in a joint 3D and 2D space. Moreover, among all compared methods, ZeroScene is the only one that supports separating objects from the scene and specifically processing the background, further enhancing its overall performance and usability. Table \ref{tab:runtime} provides an analysis of the average runtime of ZeroScene for 3D scene generation from a single image.

\noindent\textbf{\textit{Ablation study.}} We conducted ablation studies on the foreground object inpainting within our framework, and the different combinations of loss functions in Eq. \ref{eq:1} to validate their effectiveness. As illustrated in Fig. \ref{fig:ablation qualitative1} and Tab. \ref{tab:ablation quantitative1}, directly utilizing unrefined segmented images for generation leads to structural inaccuracies in the 3D assets, significantly degrading geometric metrics and impairing visual quality. When employing only $\mathcal{L}_{CD}^{2D}$ as the loss function, the lack of constraints from 3D spatial depth information results in severe positional misalignment of objects during optimization, which causes notable deviations in scene-level metrics compared to the reference data. Conversely, when solely using $\mathcal{L}_{CD}^{3D}$, while the depth prior aids in capturing relative spatial relationships between objects, the optimization of object positions and rotation parameters still exhibits convergence instability, leading to a certain degree of model degradation.

\begin{figure*}
    \centering
    \vspace{-0.3cm}
    \includegraphics[width=\textwidth]{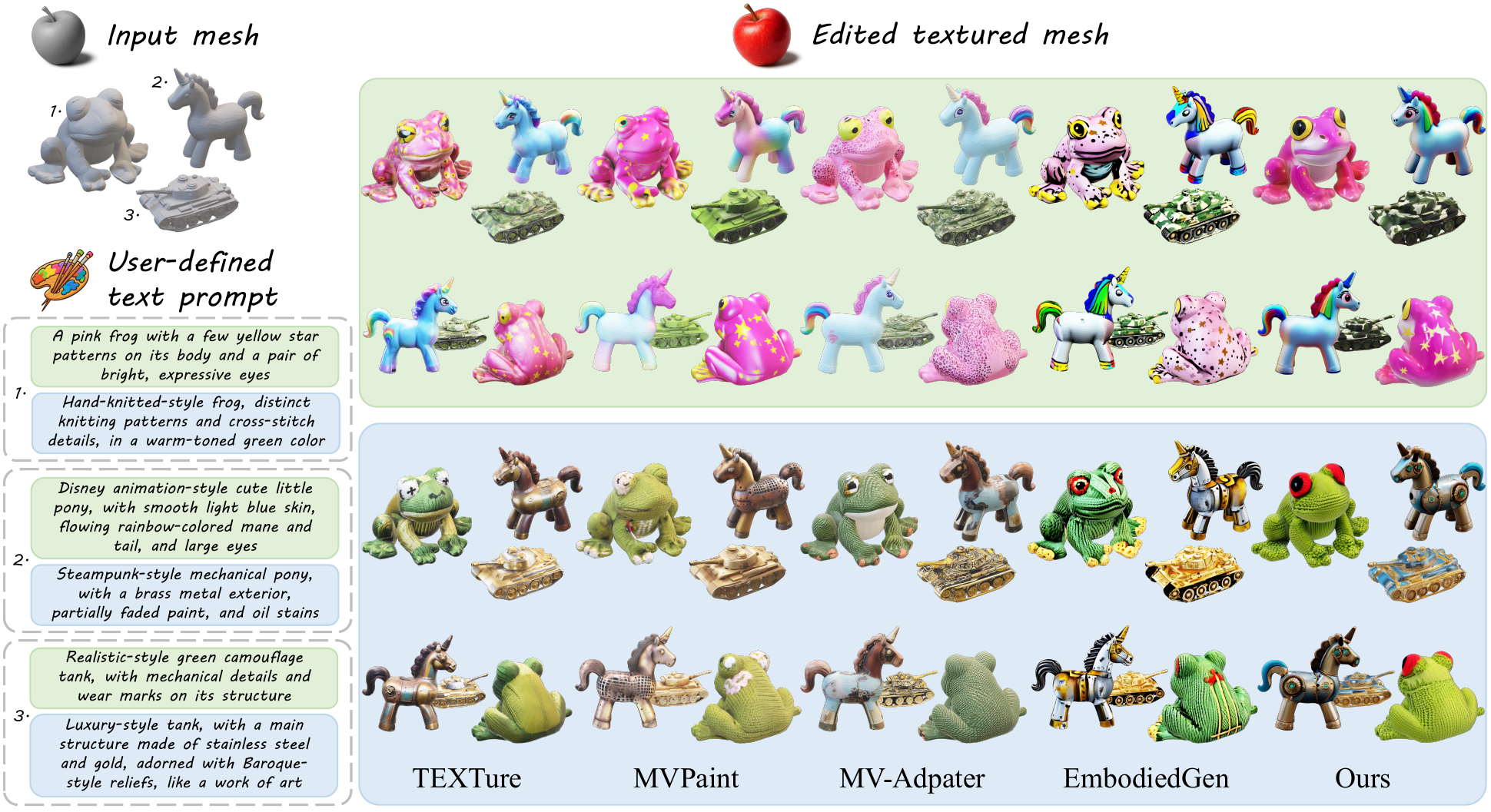}
    \vspace{-0.6cm}
    \caption{\textbf{Qualitative comparisons between ZeroScene and state-of-the-art text-to-texture generation methods.} The test scene is generated by ZeroScene, with all instances numbered and annotated with two sets of prompts (green and blue). After obtaining the texture results from each method, we reassemble the individual assets back into the scene according to the layout optimization described in Sec. \ref{sec:3.1}.}
    \vspace{-0.8cm}
    \label{fig:qualitative comparison2}
\end{figure*}

\subsection{Texture Editing} \label{sec:4.3}

\noindent\textbf{\textit{Baseline \& Benchmark.}} We compare our method against state-of-the-art texture generation approaches, including TEXTure \cite{richardson2023texture}, MVPaint \cite{cheng2025mvpaint}, MV-Adapter \cite{huang2025mv}, and EmbodiedGen \cite{wang2025embodiedgen}. To comprehensively evaluate the generalization capability of text-to-texture generation methods across various mesh structures, we construct a diverse and compact benchmark. The models in this benchmark are randomly sampled from 3D datasets (i.e., Objaverse-XL \cite{deitke2023objaverse}, GSO \cite{downs2022google}), with additional scene assets generated by ZeroScene.

\begin{figure*}
    \centering
    %\vspace{-0.3cm}
    \includegraphics[width=\textwidth]{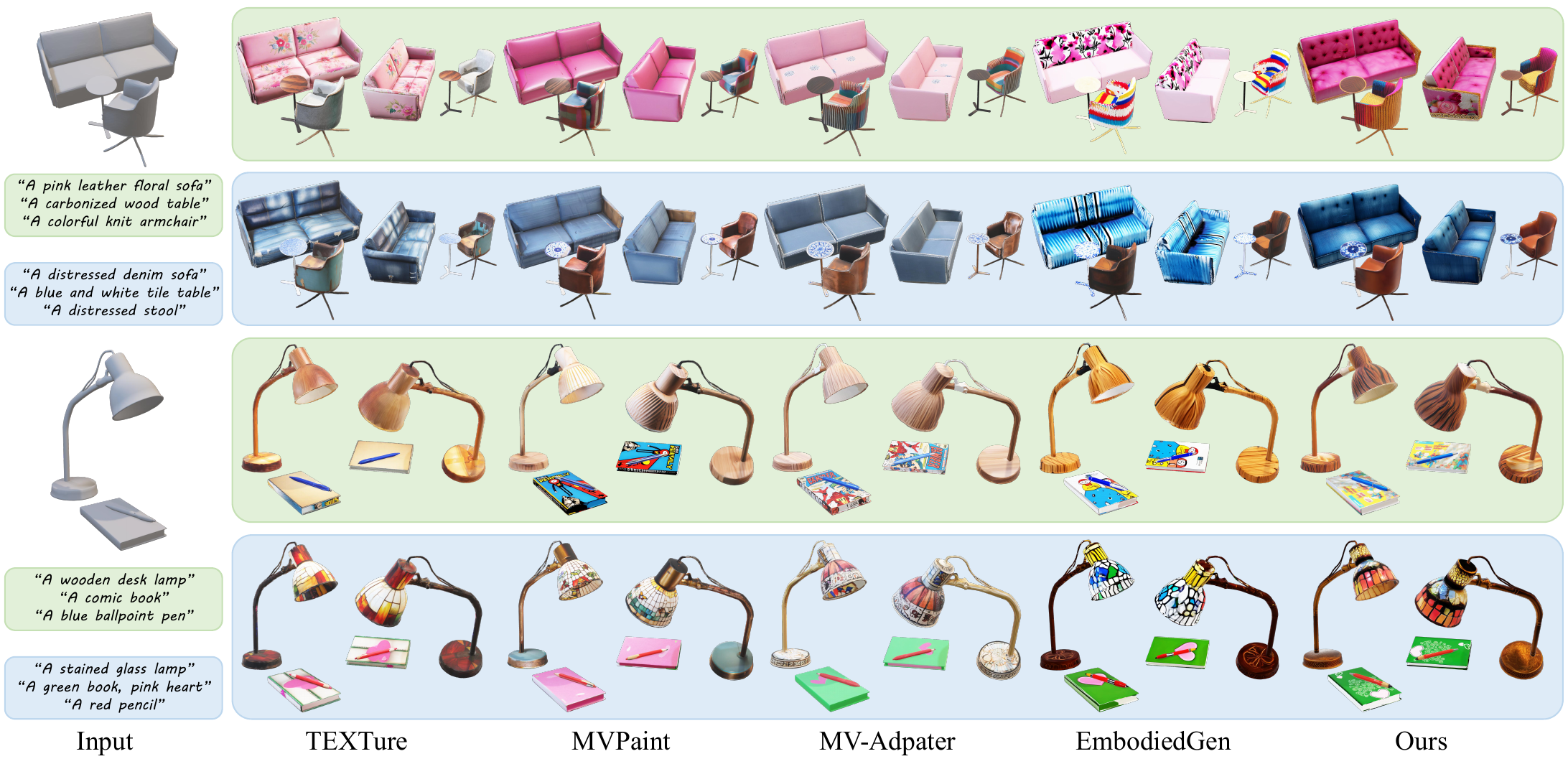}
    %\vspace{-0.7cm}
    \caption{\textbf{Additional qualitative comparison results of texture editing on generated assets.}}
    %\vspace{-0.6cm}
    \label{fig:qualitative comparison3}
\end{figure*}

\begin{figure*}[htbp]
    \centering
    %\vspace{-0.3cm}
    \includegraphics[width=\textwidth]{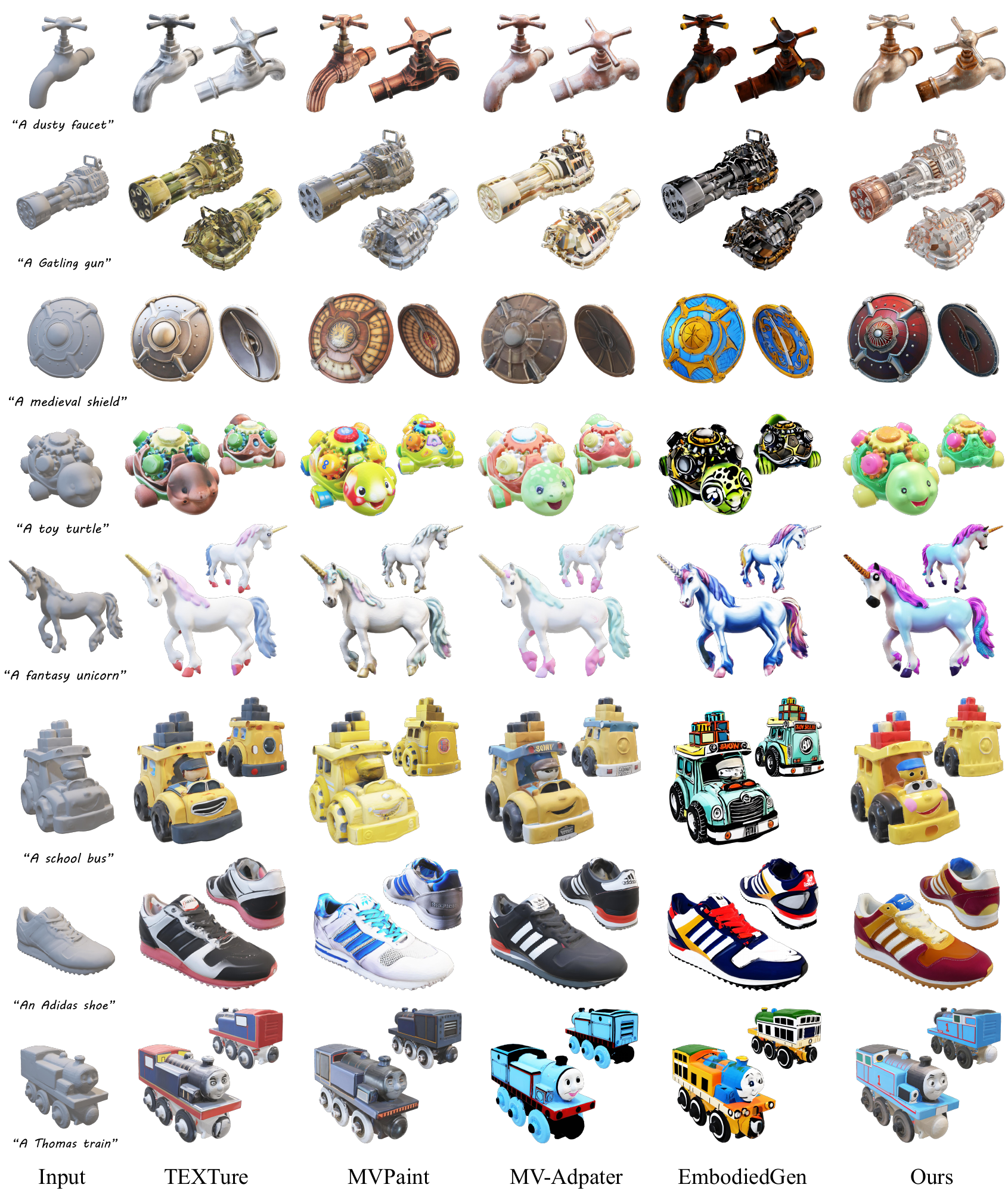}
    %\vspace{-0.7cm}
    \caption{\textbf{Additional qualitative comparison results of texture editing on the Objaverse-XL \cite{deitke2023objaverse} and GSO \cite{downs2022google} datasets.}}
    %\vspace{-0.6cm}
    \label{fig:qualitative comparison4}
\end{figure*}

\noindent\textbf{\textit{Qualitative comparison.}} Fig. \ref{fig:qualitative comparison2} presents a comparative visualization of texture synthesis results on the same mesh model under identical textual prompts, generated by various methods. As demonstrated in the results, TEXTure tends to generate noticeable seam artifacts and texture stretching at view transition regions when handling complex geometric topologies. MVPaint struggles to adequately capture key semantic information from the text and exhibits noticeable inconsistencies across different viewpoints. MV-Adapter employs a multi-view diffusion model to generate textures for textureless meshes; however, its output images often retain illumination interference, leading to visible artifacts after texture mapping. The underlying image generation model used in EmbodiedGen occasionally fails to produce multi-view content that aligns closely with the given text and is prone to distortion during back projection onto the mesh surface. In contrast, our method generates textures with high clarity and rich details that accurately reflect the visual descriptions in the prompt, while also demonstrating excellent multi-view consistency. Furthermore, by incorporating PBR materials, the textured mesh exhibits enhanced three dimensional depth and photorealism. Additional comparative results can be found in Fig. \ref{fig:qualitative comparison3} and Fig. \ref{fig:qualitative comparison4}.

\begin{table}
  \centering
  \caption{\textbf{Quantitative comparisons of texture generation methods across three metrics with FID, KID and CLIP similarity.}}
  \resizebox{\linewidth}{!}{
  \begin{tabular}{lccc}
    \toprule[1.5pt]
    \textbf{Method} & \textbf{FID} $\downarrow$  & \textbf{KID($\times 10^{-3}$)} $\downarrow$  & \textbf{CLIP} $\uparrow$ \\
    \midrule
    TEXTure \cite{richardson2023texture} & 57.31 & 9.76 & 0.811 \\
    MVPaint \cite{cheng2025mvpaint} & 54.77 & 10.15 & 0.805 \\
    MV-Adpater \cite{huang2025mv} & 51.48 & 8.34 & 0.817 \\
    EmbodiedGen \cite{wang2025embodiedgen}  & 63.62 & 11.97 & 0.733 \\
    \midrule
    Ours & \textbf{42.19} & \textbf{6.53} & \textbf{0.826} \\
    \bottomrule[1.5pt]
  \end{tabular}}
  \label{tab:quantitative comparison2}
  \vspace{-0.6cm}
\end{table}

\noindent\textbf{\textit{Quantitative comparison.}} Upon completing the texture generation for the 3D asset, we uniformly sample 36 viewpoints around the mesh model at a $10^\circ$ elevation angle, rendering images at a resolution of $512 \times 512$. For models with existing textures from the Objaverse-XL \cite{deitke2023objaverse} and GSO \cite{downs2022google} datasets, we follow the same sampling procedure and use their renderings as ground truth. Subsequently, we compare the distribution of the rendered image set obtained from the generated textures against that of the image set sampled from the real textures. To quantify the distribution discrepancy and visual quality between the two image sets, we adopt widely-used evaluation metrics in text-to-texture generation tasks: Fr\'echet Inception Distance (FID) \cite{heusel2017gans} and Kernel Inception Distance (KID) \cite{kid}. Additionally, we compute the CLIP similarity to further assess the semantic consistency between the generated textures and the input text prompts. As shown in Tab. \ref{tab:quantitative comparison2}, our method achieves lower FID and KID scores, indicating that the visual distribution of the generated textures is closer to that of real object textures. Meanwhile, the higher CLIP score demonstrates that our approach can produce textures that more accurately align with the user's input intent.

\begin{table}
  \centering
  \caption{\textbf{Quantitative analysis of ablation study.} "Mask-guided." denotes the mask-guided image generation strategy, and "PBR Mat." refers to the estimation of PBR materials.}
  \resizebox{\linewidth}{!}{
  \begin{tabular}{lccc}
    \toprule[1.5pt]
    \textbf{Ablation} & \textbf{FID} $\downarrow$  & \textbf{KID($\times 10^{-3}$)} $\downarrow$  & \textbf{CLIP} $\uparrow$ \\
    \midrule
    W/o Mask-guided. & 50.29 & 8.57 & 0.784 \\
    W/o PBR Mat. & 46.38 & 7.02 & 0.811 \\
    \midrule
    Ours & \textbf{42.19} & \textbf{6.53} & \textbf{0.826} \\
    \bottomrule[1.5pt]
  \end{tabular}}
  \label{tab:ablation quantitative2}
  \vspace{-0.6cm}
\end{table}

\begin{figure}
    \centering
    %\vspace{-0.3cm}
    \includegraphics[width=\columnwidth]{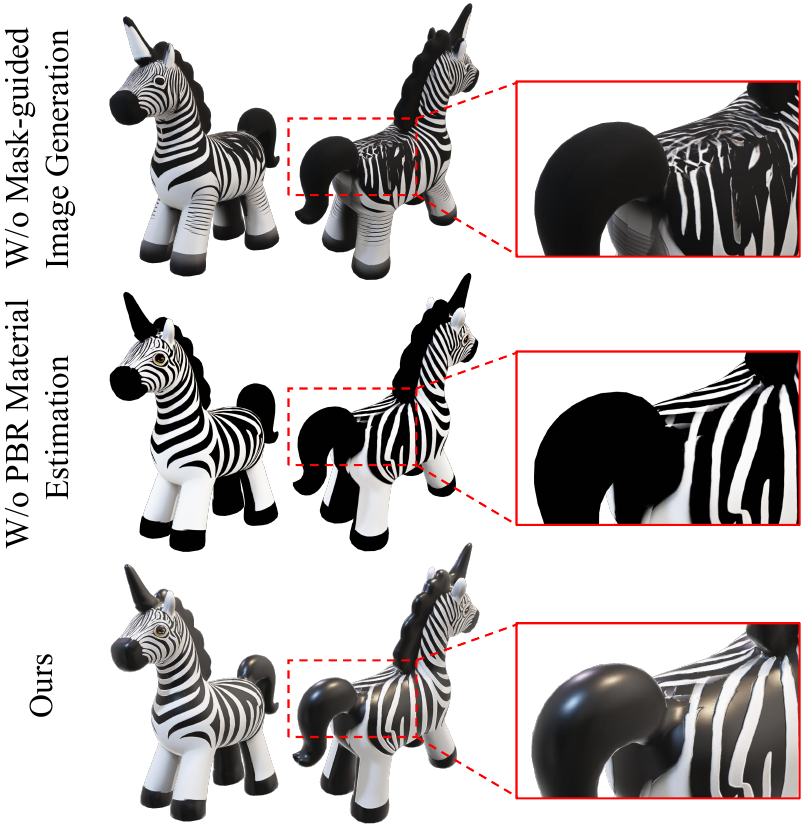}
    \vspace{-0.7cm}
    \caption{\textbf{Ablation study.} The prompt is "a plastic zebra toy". The complete design enables the generated textures to have superior consistency and rendering quality.}
    \vspace{-0.9cm}
    \label{fig:ablation qualitative2}
\end{figure}

\noindent\textbf{\textit{Ablation study.}} We conducted ablation studies on two key designs in the ZeroScene texture editing function: the mask-guided progressive image generation strategy and PBR material estimation. As shown in Fig. \ref{fig:ablation qualitative2} and Tab. \ref{tab:ablation quantitative2}, without employing the mask-guided multi-view generation approach, instead independently generating images based only on the geometric conditions of each view, it is difficult to effectively maintain consistency across multiple views. When projected onto the 3D model surface, the texture exhibits blurriness and artifacts, leading to a significant degradation in quantitative metrics. On the other hand, if PBR material estimation is omitted after obtaining the albedo map, the lack of interaction between the model and environmental lighting negatively impacts the final texture quality, diminishing the realism and three-dimensionality of the rendered results.
\section{Discussion}

\noindent\textbf{\textit{Limitations.}} In this paper, we focus on the generation of localized scenes and small-scale indoor scenes, along with texture editing of objects within such environments. In scenarios involving more complex object interactions, such as city-scale or outdoor settings, severe occlusion often leads to erroneous shape inference and further undermines the stability of layout optimization, resulting in performance degradation of current methods. At present, the types of background entities that ZeroScene can handle are relatively limited, primarily confined to supporting structures with simple geometries and clear semantics (e.g., tables, floors, and walls). It struggles to accurately generate interwoven structures commonly found in natural scenes, such as grasslands or oceans. We plan to extend our experiments to scenes with more objects and stronger occlusion relationships, and to introduce more sophisticated background handling mechanisms to enhance the system's robustness. 

Moreover, the texture editing results of ZeroScene rely on the underlying text-to-image model. Existing models still often produce images lacking in rich detail when processing certain textual prompts, particularly underperforming when representing real world objects with specific material properties (e.g., fabrics, fur, glass). Accurately modeling the material attributes of arbitrary objects remains a significant challenge in the field of 3D shape and texture generation, and constitutes a major direction for our future work.

\begin{figure}
    \centering
    %\vspace{-0.3cm}
    \includegraphics[width=\columnwidth]{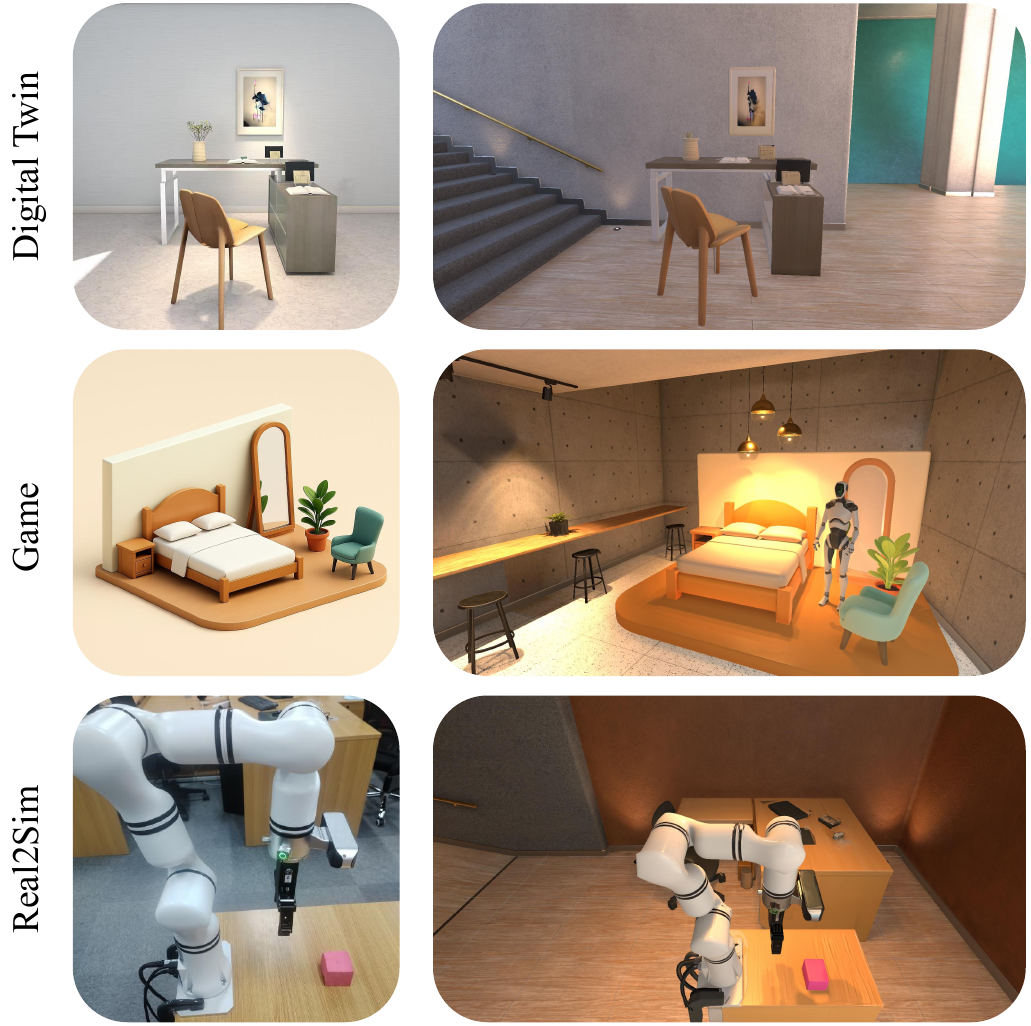}
    \vspace{-0.6cm}
    \caption{ZeroScene has demonstrated significant application value and broad prospects in numerous domains, such as generating digital twin scenes consistent with images, efficiently producing diverse game assets, and constructing virtual environments suitable for embodied intelligence simulation and training.}
    \vspace{-0.8cm}
    \label{fig:application}
\end{figure}

\noindent\textbf{\textit{Applications.}} As shown in Fig. \ref{fig:application}, ZeroScene is capable of converting a single image into a high-quality 3D scene, demonstrating broad application potential in various fields such as digital twins and real-to-simulation workflows in robotics. The generated scene assets feature an explicit mesh representation, enabling seamless integration into mainstream game development environments like Unity and Unreal Engine. Thereby, it alleviates the high costs traditionally associated with manual modeling and 3D scanning techniques to a certain extent. Furthermore, the texture editing capabilities provided by ZeroScene allow users to rapidly generate diverse 3D assets, facilitating the efficient construction of visually rich and vibrant virtual worlds.
\section{Conclusion}

In this paper, we introduce ZeroScene, a zero-shot framework capable of generating consistent 3D scenes from a single image while supporting flexible texture editing. By leveraging techniques such as scene decomposition, image inpainting, and cross-dimensional object pose optimization, ZeroScene achieves coherent reconstruction of entire scenes with harmonious integration of foreground and background, effectively overcoming limitations of existing methods in terms of insufficient asset generation quality and weak modeling of spatial relationships between objects. Furthermore, via geometry constraints and a mask-guided progressive image generation strategy, our framework enables diverse texture editing of objects within the scene while maintaining multi-view consistency. Qualitative and quantitative experimental results demonstrate that ZeroScene significantly outperforms existing methods in both the geometric structure and visual quality of generated assets, highlighting its broad application potential in immersive digital content creation.

\section*{Acknowledgements}
This work was supported in part by the Major Key Project of
PCL (PCL2025AS216 and PCL2025AS17), the National Key R\&D Program of China (2023YFA1008500), the National Natural Science Foundation of China (NSFC) under grants U22B2035, and the Fundamental and Interdisciplinary Disciplines Breakthrough Plan of the Ministry of Education of China (JYB2025XDXM901).

% bibtex
\bibliographystyle{eg-alpha-doi} 
\bibliography{egbibsample}

\newcommand{\etalchar}[1]{$^{#1}$}
\begin{thebibliography}{\uppercase{PWMAZ24}}

\bibitem[A{\"O}E25]{Ardelean2025Gen3DSR}
\textsc{Ardelean A., {\"O}zer M., Egger B.}:
\newblock Generalizable 3d scene reconstruction via divide and conquer from a
  single view.
\newblock In \emph{International Conference on 3D Vision (3DV)} (2025).

\bibitem[BBB{\etalchar{*}}25]{batifol2025flux}
\textsc{Batifol S., Blattmann A., Boesel F., Consul S., Diagne C., Dockhorn T.,
  English J., English Z., Esser P., Kulal S., et~al.}:
\newblock Flux. 1 kontext: Flow matching for in-context image generation and
  editing in latent space.
\newblock \emph{arXiv e-prints} (2025), arXiv--2506.

\bibitem[BKA{\etalchar{*}}24]{bensadoun2024meta}
\textsc{Bensadoun R., Kleiman Y., Azuri I., Harosh O., Vedaldi A., Neverova N.,
  Gafni O.}:
\newblock Meta 3d texturegen: Fast and consistent texture generation for 3d
  objects.
\newblock \emph{arXiv preprint arXiv:2407.02430} (2024).

\bibitem[BSAG18]{kid}
\textsc{Binkowski M., Sutherland D.~J., Arbel M., Gretton A.}:
\newblock Demystifying {MMD} gans.
\newblock In \emph{6th International Conference on Learning Representations}
  (2018).

\bibitem[CHLZ21]{chen2021unsupervised}
\textsc{Chen C., Han Z., Liu Y.-S., Zwicker M.}:
\newblock Unsupervised learning of fine structure generation for 3d point
  clouds by 2d projections matching.
\newblock In \emph{Proceedings of the ieee/cvf international conference on
  computer vision} (2021), pp.~12466--12477.

\bibitem[CJL{\etalchar{*}}25]{chen2025physgen3d}
\textsc{Chen B., Jiang H., Liu S., Gupta S., Li Y., Zhao H., Wang S.}:
\newblock Physgen3d: Crafting a miniature interactive world from a single
  image.
\newblock In \emph{Proceedings of the Computer Vision and Pattern Recognition
  Conference} (2025), pp.~6178--6189.

\bibitem[CMZ{\etalchar{*}}25]{cheng2025mvpaint}
\textsc{Cheng W., Mu J., Zeng X., Chen X., Pang A., Zhang C., Wang Z., Fu B.,
  Yu G., Liu Z., et~al.}:
\newblock Mvpaint: Synchronized multi-view diffusion for painting anything 3d.
\newblock In \emph{Proceedings of the Computer Vision and Pattern Recognition
  Conference} (2025), pp.~585--594.

\bibitem[CSL{\etalchar{*}}23]{chen2023text2tex}
\textsc{Chen D.~Z., Siddiqui Y., Lee H.-Y., Tulyakov S., Nie{\ss}ner M.}:
\newblock Text2tex: Text-driven texture synthesis via diffusion models.
\newblock In \emph{Proceedings of the IEEE/CVF international conference on
  computer vision} (2023), pp.~18558--18568.

\bibitem[CWS{\etalchar{*}}25]{chen2025autopartgen}
\textsc{Chen M., Wang J., Shapovalov R., Monnier T., Jung H., Wang D., Ranjan
  R., Laina I., Vedaldi A.}:
\newblock Autopartgen: Autogressive 3d part generation and discovery.
\newblock \emph{arXiv preprint arXiv:2507.13346} (2025).

\bibitem[CWW{\etalchar{*}}24]{chen2024comboverse}
\textsc{Chen Y., Wang T., Wu T., Pan X., Jia K., Liu Z.}:
\newblock Comboverse: Compositional 3d assets creation using spatially-aware
  diffusion guidance.
\newblock In \emph{European Conference on Computer Vision} (2024), Springer,
  pp.~128--146.

\bibitem[DFK{\etalchar{*}}22]{downs2022google}
\textsc{Downs L., Francis A., Koenig N., Kinman B., Hickman R., Reymann K.,
  McHugh T.~B., Vanhoucke V.}:
\newblock Google scanned objects: A high-quality dataset of 3d scanned
  household items.
\newblock In \emph{2022 International Conference on Robotics and Automation
  (ICRA)} (2022), IEEE, pp.~2553--2560.

\bibitem[DLW{\etalchar{*}}23]{deitke2023objaverse}
\textsc{Deitke M., Liu R., Wallingford M., Ngo H., Michel O., Kusupati A., Fan
  A., Laforte C., Voleti V., Gadre S.~Y., et~al.}:
\newblock Objaverse-xl: A universe of 10m+ 3d objects.
\newblock \emph{Advances in Neural Information Processing Systems 36} (2023),
  35799--35813.

\bibitem[DWJ{\etalchar{*}}24]{daiautomated}
\textsc{Dai T., Wong J., Jiang Y., Wang C., Gokmen C., Zhang R., Wu J., Fei-Fei
  L.}:
\newblock Automated creation of digital cousins for robust policy learning.
\newblock In \emph{8th Annual Conference on Robot Learning} (2024).

\bibitem[DYY{\etalchar{*}}25]{dong2025hiscene}
\textsc{Dong W., Yang B., Yang Z., Li Y., Hu T., Bao H., Ma Y., Cui Z.}:
\newblock Hiscene: creating hierarchical 3d scenes with isometric view
  generation.
\newblock In \emph{Proceedings of the 33rd ACM International Conference on
  Multimedia} (2025), pp.~9783--9792.

\bibitem[FCG{\etalchar{*}}21]{fu20213d}
\textsc{Fu H., Cai B., Gao L., Zhang L.-X., Wang J., Li C., Zeng Q., Sun C.,
  Jia R., Zhao B., et~al.}:
\newblock 3d-front: 3d furnished rooms with layouts and semantics.
\newblock In \emph{Proceedings of the IEEE/CVF International Conference on
  Computer Vision} (2021), pp.~10933--10942.

\bibitem[FYY{\etalchar{*}}25]{feng2025romantex}
\textsc{Feng Y., Yang M., Yang S., Zhang S., Yu J., Zhao Z., Liu Y., Jiang J.,
  Guo C.}:
\newblock Romantex: Decoupling 3d-aware rotary positional embedded
  multi-attention network for texture synthesis.
\newblock In \emph{Proceedings of the IEEE/CVF international conference on
  computer vision} (2025).

\bibitem[GCL{\etalchar{*}}25]{gu2025artiscene}
\textsc{Gu Z., Cui Y., Li Z., Wei F., Ge Y., Gu J., Liu M.-Y., Davis A., Ding
  Y.}:
\newblock Artiscene: Language-driven artistic 3d scene generation through image
  intermediary.
\newblock In \emph{Proceedings of the Computer Vision and Pattern Recognition
  Conference} (2025), pp.~2891--2901.

\bibitem[GPAM{\etalchar{*}}14]{goodfellow2014generative}
\textsc{Goodfellow I.~J., Pouget-Abadie J., Mirza M., Xu B., Warde-Farley D.,
  Ozair S., Courville A., Bengio Y.}:
\newblock Generative adversarial nets.
\newblock \emph{Advances in neural information processing systems 27} (2014).

\bibitem[HGA{\etalchar{*}}25]{huang2025midi}
\textsc{Huang Z., Guo Y.-C., An X., Yang Y., Li Y., Zou Z.-X., Liang D., Liu
  X., Cao Y.-P., Sheng L.}:
\newblock Midi: Multi-instance diffusion for single image to 3d scene
  generation.
\newblock In \emph{Proceedings of the Computer Vision and Pattern Recognition
  Conference} (2025), pp.~23646--23657.

\bibitem[HGW{\etalchar{*}}25]{huang2025mv}
\textsc{Huang Z., Guo Y.-C., Wang H., Yi R., Ma L., Cao Y.-P., Sheng L.}:
\newblock Mv-adapter: Multi-view consistent image generation made easy.
\newblock In \emph{Proceedings of the IEEE/CVF International Conference on
  Computer Vision} (2025), pp.~16377--16387.

\bibitem[HGZ{\etalchar{*}}24]{huo2024texgen}
\textsc{Huo D., Guo Z., Zuo X., Shi Z., Lu J., Dai P., Xu S., Cheng L., Yang
  Y.-H.}:
\newblock Texgen: Text-guided 3d texture generation with multi-view sampling
  and resampling.
\newblock In \emph{European Conference on Computer Vision} (2024), Springer,
  pp.~352--368.

\bibitem[HJA20]{ho2020denoising}
\textsc{Ho J., Jain A., Abbeel P.}:
\newblock Denoising diffusion probabilistic models.
\newblock \emph{Advances in neural information processing systems 33} (2020),
  6840--6851.

\bibitem[HLG{\etalchar{*}}24]{hurst2024gpt}
\textsc{Hurst A., Lerer A., Goucher A.~P., Perelman A., Ramesh A., Clark A.,
  Ostrow A., Welihinda A., Hayes A., Radford A., et~al.}:
\newblock Gpt-4o system card.
\newblock \emph{arXiv preprint arXiv:2410.21276} (2024).

\bibitem[HRU{\etalchar{*}}17]{heusel2017gans}
\textsc{Heusel M., Ramsauer H., Unterthiner T., Nessler B., Hochreiter S.}:
\newblock Gans trained by a two time-scale update rule converge to a local nash
  equilibrium.
\newblock \emph{Advances in neural information processing systems 30} (2017).

\bibitem[HWLW25]{huang2025material}
\textsc{Huang X., Wang T., Liu Z., Wang Q.}:
\newblock Material anything: Generating materials for any 3d object via
  diffusion.
\newblock In \emph{Proceedings of the Computer Vision and Pattern Recognition
  Conference} (2025), pp.~26556--26565.

\bibitem[HYL{\etalchar{*}}25]{han2025reparo}
\textsc{Han H., Yang R., Liao H., Xing J., Xu Z., Yu X., Zha J., Li X., Li W.}:
\newblock Reparo: Compositional 3d assets generation with differentiable 3d
  layout alignment.
\newblock In \emph{Proceedings of the IEEE/CVF International Conference on
  Computer Vision} (2025), pp.~25367--25377.

\bibitem[HYY{\etalchar{*}}25]{he2025materialmvp}
\textsc{He Z., Yang M., Yang S., Tang Y., Wang T., Zhang K., Chen G., Liu Y.,
  Jiang J., Guo C., et~al.}:
\newblock Materialmvp: Illumination-invariant material generation via
  multi-view pbr diffusion.
\newblock In \emph{Proceedings of the IEEE/CVF International Conference on
  Computer Vision} (2025).

\bibitem[KMR{\etalchar{*}}23]{kirillov2023segment}
\textsc{Kirillov A., Mintun E., Ravi N., Mao H., Rolland C., Gustafson L., Xiao
  T., Whitehead S., Berg A.~C., Lo W.-Y., et~al.}:
\newblock Segment anything.
\newblock In \emph{Proceedings of the IEEE/CVF international conference on
  computer vision} (2023), pp.~4015--4026.

\bibitem[LXLW24]{liu2024text}
\textsc{Liu Y., Xie M., Liu H., Wong T.-T.}:
\newblock Text-guided texturing by synchronized multi-view diffusion.
\newblock In \emph{SIGGRAPH Asia 2024 Conference Papers} (2024), pp.~1--11.

\bibitem[LZL{\etalchar{*}}25a]{lai2025hunyuan3d}
\textsc{Lai Z., Zhao Y., Liu H., Zhao Z., Lin Q., Shi H., Yang X., Yang M.,
  Yang S., Feng Y., et~al.}:
\newblock Hunyuan3d 2.5: Towards high-fidelity 3d assets generation with
  ultimate details.
\newblock \emph{arXiv preprint arXiv:2506.16504} (2025).

\bibitem[LZL{\etalchar{*}}25b]{li2025triposg}
\textsc{Li Y., Zou Z.-X., Liu Z., Wang D., Liang Y., Yu Z., Liu X., Guo Y.-C.,
  Liang D., Ouyang W., et~al.}:
\newblock Triposg: High-fidelity 3d shape synthesis using large-scale rectified
  flow models.
\newblock \emph{arXiv preprint arXiv:2502.06608} (2025).

\bibitem[LZR{\etalchar{*}}24]{liu2024grounding}
\textsc{Liu S., Zeng Z., Ren T., Li F., Zhang H., Yang J., Jiang Q., Li C.,
  Yang J., Su H., et~al.}:
\newblock Grounding dino: Marrying dino with grounded pre-training for open-set
  object detection.
\newblock In \emph{European conference on computer vision} (2024), Springer,
  pp.~38--55.

\bibitem[LZZZ24]{liu2024controllable}
\textsc{Liu J.-H., Zhang S.-K., Zhang C., Zhang S.-H.}:
\newblock Controllable procedural generation of landscapes.
\newblock In \emph{Proceedings of the 32nd ACM International Conference on
  Multimedia} (2024), pp.~6394--6403.

\bibitem[MWZX26]{meng2025scenegen}
\textsc{Meng Y., Wu H., Zhang Y., Xie W.}:
\newblock Scenegen: Single-image 3d scene generation in one feedforward pass.
\newblock In \emph{2026 International Conference on 3D Vision (3DV)} (2026).

\bibitem[ODM{\etalchar{*}}24]{oquab2024dinov2}
\textsc{Oquab M., Darcet T., Moutakanni T., Vo H., Szafraniec M., Khalidov V.,
  Fernandez P., Haziza D., Massa F., El-Nouby A., et~al.}:
\newblock Dinov2: Learning robust visual features without supervision.
\newblock \emph{Transactions on Machine Learning Research Journal} (2024).

\bibitem[PWMAZ24]{perla2024easitex}
\textsc{Perla S. R.~K., Wang Y., Mahdavi-Amiri A., Zhang H.}:
\newblock Easi-tex: Edge-aware mesh texturing from single image.
\newblock \emph{ACM Transactions on Graphics 43}, 4 (2024).

\bibitem[RBL{\etalchar{*}}22]{rombach2022high}
\textsc{Rombach R., Blattmann A., Lorenz D., Esser P., Ommer B.}:
\newblock High-resolution image synthesis with latent diffusion models.
\newblock In \emph{Proceedings of the IEEE/CVF conference on computer vision
  and pattern recognition} (2022), pp.~10684--10695.

\bibitem[RKH{\etalchar{*}}21]{radford2021learning}
\textsc{Radford A., Kim J.~W., Hallacy C., Ramesh A., Goh G., Agarwal S.,
  Sastry G., Askell A., Mishkin P., Clark J., et~al.}:
\newblock Learning transferable visual models from natural language
  supervision.
\newblock In \emph{International conference on machine learning} (2021), PmLR,
  pp.~8748--8763.

\bibitem[RLZ{\etalchar{*}}24]{ren2024grounded}
\textsc{Ren T., Liu S., Zeng A., Lin J., Li K., Cao H., Chen J., Huang X., Chen
  Y., Yan F., et~al.}:
\newblock Grounded sam: Assembling open-world models for diverse visual tasks.
\newblock \emph{arXiv preprint arXiv:2401.14159} (2024).

\bibitem[RMA{\etalchar{*}}23]{richardson2023texture}
\textsc{Richardson E., Metzer G., Alaluf Y., Giryes R., Cohen-Or D.}:
\newblock Texture: Text-guided texturing of 3d shapes.
\newblock In \emph{ACM SIGGRAPH 2023 conference proceedings} (2023), pp.~1--11.

\bibitem[SCS{\etalchar{*}}22]{saharia2022photorealistic}
\textsc{Saharia C., Chan W., Saxena S., Li L., Whang J., Denton E.~L.,
  Ghasemipour K., Gontijo~Lopes R., Karagol~Ayan B., Salimans T., et~al.}:
\newblock Photorealistic text-to-image diffusion models with deep language
  understanding.
\newblock \emph{Advances in neural information processing systems 35} (2022),
  36479--36494.

\bibitem[SLM{\etalchar{*}}22]{suvorov2022resolution}
\textsc{Suvorov R., Logacheva E., Mashikhin A., Remizova A., Ashukha A.,
  Silvestrov A., Kong N., Goka H., Park K., Lempitsky V.}:
\newblock Resolution-robust large mask inpainting with fourier convolutions.
\newblock In \emph{Proceedings of the IEEE/CVF winter conference on
  applications of computer vision} (2022), pp.~2149--2159.

\bibitem[Tea24]{team2024kolors}
\textsc{Team K.}:
\newblock Kolors: Effective training of diffusion model for photorealistic
  text-to-image synthesis.
\newblock \emph{arXiv preprint} (2024).

\bibitem[Tea25]{hunyuan3d22025tencent}
\textsc{Team T.~H.}:
\newblock Hunyuan3d 2.0: Scaling diffusion models for high resolution textured
  3d assets generation, 2025.

\bibitem[WIR{\etalchar{*}}25]{wu2025diorama}
\textsc{Wu Q., Iliash D., Ritchie D., Savva M., Chang A.~X.}:
\newblock Diorama: Unleashing zero-shot single-view 3d indoor scene modeling.
\newblock In \emph{Proceedings of the IEEE/CVF International Conference on
  Computer Vision} (2025), pp.~8896--8907.

\bibitem[WLC{\etalchar{*}}24a]{wang2024dust3r}
\textsc{Wang S., Leroy V., Cabon Y., Chidlovskii B., Revaud J.}:
\newblock Dust3r: Geometric 3d vision made easy.
\newblock In \emph{Proceedings of the IEEE/CVF Conference on Computer Vision
  and Pattern Recognition} (2024), pp.~20697--20709.

\bibitem[WLC{\etalchar{*}}24b]{wu2024unique3d}
\textsc{Wu K., Liu F., Cai Z., Yan R., Wang H., Hu Y., Duan Y., Ma K.}:
\newblock Unique3d: High-quality and efficient 3d mesh generation from a single
  image.
\newblock \emph{Advances in Neural Information Processing Systems 37} (2024),
  125116--125141.

\bibitem[WLC{\etalchar{*}}25]{wang2025embodiedgen}
\textsc{Wang X., Liu L., Cao Y., Wu R., Qin W., Wang D., Sui W., Su Z.}:
\newblock Embodiedgen: Towards a generative 3d world engine for embodied
  intelligence.
\newblock \emph{arXiv preprint arXiv:2506.10600} (2025).

\bibitem[WLZ{\etalchar{*}}25]{wu2025direct3d}
\textsc{Wu S., Lin Y., Zhang F., Zeng Y., Yang Y., Bao Y., Qian J., Zhu S., Cao
  X., Torr P., et~al.}:
\newblock Direct3d-s2: Gigascale 3d generation made easy with spatial sparse
  attention.
\newblock \emph{Advances in Neural Information Processing Systems} (2025).

\bibitem[WXDS21]{wang2021real}
\textsc{Wang X., Xie L., Dong C., Shan Y.}:
\newblock Real-esrgan: Training real-world blind super-resolution with pure
  synthetic data.
\newblock In \emph{Proceedings of the IEEE/CVF international conference on
  computer vision} (2021), pp.~1905--1914.

\bibitem[XGF{\etalchar{*}}25]{xiang2025make}
\textsc{Xiang X., Gorelik L.~S., Fan Y., Armstrong O., Iandola F., Li Y.,
  Lifshitz I., Ranjan R.}:
\newblock Make-a-texture: Fast shape-aware texture generation in 3 seconds.
\newblock In \emph{2025 IEEE/CVF Winter Conference on Applications of Computer
  Vision (WACV)} (2025), IEEE, pp.~4872--4881.

\bibitem[XLX{\etalchar{*}}25]{xiang2025structured}
\textsc{Xiang J., Lv Z., Xu S., Deng Y., Wang R., Zhang B., Chen D., Tong X.,
  Yang J.}:
\newblock Structured 3d latents for scalable and versatile 3d generation.
\newblock In \emph{Proceedings of the Computer Vision and Pattern Recognition
  Conference} (2025), pp.~21469--21480.

\bibitem[YOPM24]{youwang2024paint}
\textsc{Youwang K., Oh T.-H., Pons-Moll G.}:
\newblock Paint-it: Text-to-texture synthesis via deep convolutional texture
  map optimization and physically-based rendering.
\newblock In \emph{Proceedings of the ieee/cvf conference on computer vision
  and pattern recognition} (2024), pp.~4347--4356.

\bibitem[YSL{\etalchar{*}}25]{yang2025fast3r}
\textsc{Yang J., Sax A., Liang K.~J., Henaff M., Tang H., Cao A., Chai J.,
  Meier F., Feiszli M.}:
\newblock Fast3r: Towards 3d reconstruction of 1000+ images in one forward
  pass.
\newblock In \emph{Proceedings of the Computer Vision and Pattern Recognition
  Conference} (2025), pp.~21924--21935.

\bibitem[YWL{\etalchar{*}}25]{ye2025hi3dgen}
\textsc{Ye C., Wu Y., Lu Z., Chang J., Guo X., Zhou J., Zhao H., Han X.}:
\newblock Hi3dgen: High-fidelity 3d geometry generation from images via normal
  bridging.
\newblock In \emph{Proceedings of the IEEE/CVF International Conference on
  Computer Vision} (2025).

\bibitem[YZY{\etalchar{*}}25]{yao2025cast}
\textsc{Yao K., Zhang L., Yan X., Zeng Y., Zhang Q., Xu L., Yang W., Gu J., Yu
  J.}:
\newblock Cast: Component-aligned 3d scene reconstruction from an rgb image.
\newblock \emph{ACM Transactions on Graphics (TOG) 44}, 4 (2025), 1--19.

\bibitem[ZCQ{\etalchar{*}}24]{zeng2024paint3d}
\textsc{Zeng X., Chen X., Qi Z., Liu W., Zhao Z., Wang Z., Fu B., Liu Y., Yu
  G.}:
\newblock Paint3d: Paint anything 3d with lighting-less texture diffusion
  models.
\newblock In \emph{Proceedings of the IEEE/CVF conference on computer vision
  and pattern recognition} (2024), pp.~4252--4262.

\bibitem[ZLH24]{zhou2024zero}
\textsc{Zhou J., Liu Y.-S., Han Z.}:
\newblock Zero-shot scene reconstruction from single images with deep prior
  assembly.
\newblock \emph{Advances in Neural Information Processing Systems 37} (2024),
  39104--39127.

\bibitem[ZPX{\etalchar{*}}24]{zhang2024mapa}
\textsc{Zhang S., Peng S., Xu T., Yang Y., Chen T., Xue N., Shen Y., Bao H., Hu
  R., Zhou X.}:
\newblock Mapa: Text-driven photorealistic material painting for 3d shapes.
\newblock In \emph{ACM SIGGRAPH 2024 Conference Papers} (2024), pp.~1--12.

\bibitem[ZRA23]{zhang2023adding}
\textsc{Zhang L., Rao A., Agrawala M.}:
\newblock Adding conditional control to text-to-image diffusion models.
\newblock In \emph{Proceedings of the IEEE/CVF international conference on
  computer vision} (2023), pp.~3836--3847.

\bibitem[ZWZ{\etalchar{*}}24]{zhang2024clay}
\textsc{Zhang L., Wang Z., Zhang Q., Qiu Q., Pang A., Jiang H., Yang W., Xu L.,
  Yu J.}:
\newblock Clay: A controllable large-scale generative model for creating
  high-quality 3d assets.
\newblock \emph{ACM Transactions on Graphics (TOG) 43}, 4 (2024), 1--20.

\end{thebibliography}

\end{document}